\documentclass[12pt,preprint]{aastex}

\usepackage{graphicx,graphics}
\usepackage{amsmath}

\usepackage{natbib}   

\shorttitle{AMiBA Hexapod}
\shortauthors{Koch et al.}

\begin{document}

%% LaTeX will automatically break titles if they run longer than
%% one line. However, you may use \\ to force a line break if
%% you desire.

\title{The AMiBA Hexapod Telescope Mount}

%% Use \author, \affil, and the \and command to format
%% author and affiliation information.
%% Note that \email has replaced the old \authoremail command
%% from AASTeX v4.0. You can use \email to mark an email address
%% anywhere in the paper, not just in the front matter.
%% As in the title, use \\ to force line breaks.

\author{
Patrick M. Koch\altaffilmark{1},
Michael Kesteven\altaffilmark{2},
Hiroaki Nishioka\altaffilmark{1},
Homin Jiang\altaffilmark{1},
Kai-Yang Lin\altaffilmark{1,3},
Keiichi Umetsu\altaffilmark{1,4},
Yau-De Huang\altaffilmark{1},
Philippe Raffin\altaffilmark{1},
Ke-Jung Chen\altaffilmark{1},
Fabiola Iba\~nez-Romano\altaffilmark{1},
Guillaume Chereau\altaffilmark{1},
Chih-Wei Locutus Huang\altaffilmark{3,4},
Ming-Tang Chen\altaffilmark{1}, 
Paul T. P. Ho\altaffilmark{1,5},
Konrad Pausch\altaffilmark{6},
Klaus Willmeroth\altaffilmark{6},
Pablo Altamirano\altaffilmark{1},
Chia-Hao Chang\altaffilmark{1},
Shu-Hao Chang\altaffilmark{1},
Su-Wei Chang\altaffilmark{1},
Chih-Chiang Han\altaffilmark{1},
Derek Kubo\altaffilmark{1}, 
Chao-Te Li\altaffilmark{1},
Yu-Wei Liao\altaffilmark{3,4},
Guo-Chin Liu\altaffilmark{1,7},
Pierre Martin-Cocher\altaffilmark{1},
Peter Oshiro\altaffilmark{1},
Fu-Cheng Wang\altaffilmark{3,4},
Ta-Shun Wei\altaffilmark{1},
Jiun-Huei Proty Wu\altaffilmark{3,4},
Mark Birkinshaw\altaffilmark{8},
Tzihong Chiueh\altaffilmark{3},
Katy Lancaster\altaffilmark{8},
Kwok Yung Lo\altaffilmark{9},
Robert N.Martin\altaffilmark{10},
Sandor M. Molnar\altaffilmark{1},
Ferdinand Patt\altaffilmark{11} \&
Bob Romeo\altaffilmark{10}
}

\altaffiltext{1}{Academia Sinica, Institute of Astronomy and Astrophysics,
P.O.Box 23-141, Taipei 10617, Taiwan}
\altaffiltext{2}{Australia Telescope National Facility, P.O.Box 76, Epping NSW
1710, Australia}
\altaffiltext{3}{Department of Physics, Institute of Astrophysics, \& Center
for Theoretical Sciences, National Taiwan University, Taipei 10617, Taiwan}
\altaffiltext{4}{Leung center for Cosmology and Particle Astrophysics, 
National Taiwan University, Taipei 10617, Taiwan}
\altaffiltext{5}{Harvard-Smithsonian Center for Astrophysics, 60 Garden Street, Cambridge, 
MA 02138, USA}
\altaffiltext{6}{Vertex Antennentechnik GmbH, Duisburg, Germany}
\altaffiltext{7}{Department of Physics, Tamkang University, 251-37
 Tamsui, Taipei County, Taiwan} 
\altaffiltext{8}{Department of Physics, University of Bristol, Tyndall Ave, Bristol, BS8 1TL, UK}
\altaffiltext{9}{National Radio Astronomy Observatory, 520 Edgemont Road, Charlottesville, VA 22903, USA}
\altaffiltext{10}{CMA, 1638 S.Research Loop \#100, Tucson, AZ 85710, USA}
\altaffiltext{11}{ESO Headquarters Garching, Germany}

\email{pmkoch@asiaa.sinica.edu.tw}

%% Mark off your abstract in the ``abstract'' environment. In the manuscript
%% style, abstract will output a Received/Accepted line after the
%% title and affiliation information. No date will appear since the author
%% does not have this information. The dates will be filled in by the
%% editorial office after submission.

\begin{abstract}
AMiBA is the largest hexapod astronomical telescope in current
operation. We present a description of this novel hexapod mount with its
main mechanical components --- the support cone, universal joints,
jack screws, and platform --- and outline the control system with
the pointing model and the operating modes that are supported. 
The AMiBA hexapod mount performance is verified based on optical pointing tests and 
platform photogrammetry measurements.
The photogrammetry results
show that the deformations in the inner part of the platform are less than $120\mu$m rms.
This is negligible for optical pointing corrections, radio alignment and radio phase 
errors for the currently operational 7-element compact configuration. 
The optical pointing error in azimuth and elevation 
is successively reduced by a series of corrections to about 0.4$^{\prime}$ rms which meets our goal 
for the 7-element target specifications.
\end{abstract}

%% Keywords should appear after the \end{abstract} command. The uncommented
%% example has been keyed in ApJ style. See the instructions to authors
%% for the journal to which you are submitting your paper to determine
%% what keyword punctuation is appropriate.


%% From the front matter, we move on to the body of the paper.
%% In the first two sections, notice the use of the natbib \citep
%% and \citet commands to identify citations.  The citations are
%% tied to the reference list via symbolic KEYs. The KEY corresponds
%% to the KEY in the \bibitem in the reference list below. We have
%% chosen the first three characters of the first author's name plus
%% the last two numeral of the year of publication as our KEY for
%% each reference.

%% Authors who wish to have the most important objects in their paper
%% linked in the electronic edition to a data center may do so by tagging
%% their objects with \objectname{} or \object{}.  Each macro takes the
%% object name as its required argument. The optional, square-bracket 
%% argument should be used in cases where the data center identification
%% differs from what is to be printed in the paper.  The text appearing 
%% in curly braces is what will appear in print in the published paper. 
%% If the object name is recognized by the data centers, it will be linked
%% in the electronic edition to the object data available at the data centers  
%%
%% Note that for sources with brackets in their names, e.g. [WEG2004] 14h-090,
%% the brackets must be escaped with backslashes when used in the first
%% square-bracket argument, for instance, \object[\[WEG2004\] 14h-090]{90}).
%%  Otherwise, LaTeX will issue an error. 

\section{Introduction}
 
The Array for Microwave Background Anisotropy (AMiBA) is 
a dual-channel 86-102 GHz interferometer array of up to 19 elements
with a resolution up to 2$^{\prime}$.
The AMiBA --- located at the Mauna Loa weather station at 
an elevation of 3.400~m on Big Island, Hawaii --- targets 
specifically the distribution of high
redshift clusters of galaxies via the Sunyaev-Zel'dovich Effect, (e.g. \citet{sz72,bi99} and references therein),
and the anisotropies in the  
Cosmic Microwave Background (CMB), (e.g. \citet{pea99}).

In the initial AMiBA operational phase seven close-packed 0.6m diameter Cassegrain antennas are
co-mounted on a fully steerable platform controlled by a hexapod mount. 
The typical system noise temperature is $\sim 100$~K. From our observations we estimate 
a sensitivity of $\sim$ 60~mJy in 1 hour.
Previous progress reports were given in \citet{li06} and \citet{raffin06}.
The project as a whole, the correlator and the receivers are described elsewhere \citep{ho08,chen08}.
Observing strategy, calibration scheme and data analysis with quality
checks are described in \citet{lin08,wu08,nishioka08}.
First AMiBA science results are presented in \citet{huang08a,koch08b,liu08,umetsu08,wu08}.

In this paper, we describe the AMiBA mount, which is the largest
operating astronomical hexapod mount.
The role of this paper is to provide additional instrumentation details about this novel
hexapod, which complements the science papers.
Section \ref{technical} introduces the hexapod mount, with more technical details about its 
components in appendix \ref{appendix_components}. Section \ref{kinematics} gives an overview of 
the pointing corrections and the control system. The explicit pointing model is presented in 
appendix \ref{appendix_pointing}.
Photogrammetry measurements and detailed optical pointing tests verify the mount performance, 
section \ref{performance}. Our conclusions are given in section \ref{conclusion}.

\section{Hexapod Telescope} \label{technical}

The design of the AMiBA mount was driven by the requirement of having a lightweight structure which can easily and 
quickly be dismantled and shipped to another site for a possible later relocation  and the need of having 
direct access to the receivers on the 
platform for maintenance. The targeted science defined the operating frequency (86-102~GHz), and hence the required
range of baselines,
leading to a 6m platform.
Array configurations with maximum baselines require a pointing accuracy of  $\sim 0.2^{\prime}$ .
Based on these considerations a hexapod mount with a CFRP (carbon fiber reinforced plastic) platform was chosen.

Whereas the concept of the hexapod (also called Stewart platform, \citet{gough56, stewart65}) is successfully used in many technical
applications like machine tools, flight simulations or complex orthopedic surgery, its application in astronomy
has so far been mostly limited to secondary mirrors for classical Cassegrain telescopes, where the hexapod is 
used for focus optimization or wobbling movements to cancel the atmosphere and receiver noise. A pioneering 
design of a 1.5m hexapod telescope for optical astronomy was presented in \cite{chini}.
Besides this, the AMiBA is the only operating hexapod telescope. The six independently actuated
legs give the Stewart platform six degrees of freedom (x,y,z, pitch, roll and yaw), where the lengths of the legs
are changed to orient and position the platform. This parallel kinematics system has advantages and disadvantages
compared to a serial kinematics system. There is no accumulation of position errors and a generally lower inertia
allows for faster accelerations and slewing velocities. The lower mass, however brings some risk for oscillations.
The control system for the six legs is more complex,
because of more degrees of freedom in motions which can compete and interfere with each other.

For astronomical applications the hexapod offers some interesting possibilities: 
 no elevation counterweights and no azimuth bearing are needed and there is no zenith keyhole compared
to a conventional mount. Access
to the receivers and correlator from beneath the platform is straightforward.
The sky field rotation (see appendix \ref{appendix_angle})
can be compensated, and polarization measurements are possible by
rotating the entire platform.

The AMiBA hexapod with its local control system was designed and fabricated by 
Vertex Antennentechnik GmbH, Duisburg, Germany. 
After a factory acceptance test in 2004 in Germany, the whole telescope was dismantled, shipped to Hilo, 
Hawaii and assembled again on the Mauna Loa site with a final on-site acceptance test in October 2005.

The key components in the Vertex design are: the upper and lower
universal joints (u-joints) with jack screws, which require high stiffness and
large travel ranges; the stiffness of the support cone, to
minimize pointing errors from the cone; and the pointing error
model required to meet the $0^\prime\llap{.}2$ pointing
requirement.

The AMiBA hexapod has a lower limit of 30$^{\circ}$ in elevation.
Azimuth movement is unlimited without interruption. The hexapod platform polarization ($hexpol$) range 
is limited to $\pm 30^{\circ}$, with the polarization rotation defined around the pointing
axis at any possible mount position.
Both limits are chosen for safety reasons based on structural concerns. Mechanical hard
limit switches are in place to prevent movement beyond these limits in case of software failure 
and overriding.
The maximum slewing speed is 0.67$^{\circ}$/s. The telescope is 
designed to meet the harsh environmental conditions on Mauna Loa, allowing us to operate with 
 wind speeds of up to 
30 m/s (for survival in stow position, wind speeds of up to 65 m/s can be tolerated) 
and an operating temperature range of $-10^{\circ}$C to $30^{\circ}$C
(in stow position:  $-30^{\circ}$C to $30^{\circ}$C). 
Earthquake survival conditions are met with 0.3 g for both horizontal and 
vertical accelerations. The total weight of the mount is $\sim$ 31,800 kg. (support cone $\sim$ 16,600 kg,
jack screws $\sim$ 6,000 kg, universal joints $\sim$ 9,200 kg.) The platform with the interface 
ring adds another $\sim$ 3,000 kg. The current load for the 7 element system is $\sim$ 500 kg. 
The hexapod mount system, schematically illustrated in Figure \ref{fig1}, 
mainly consists of a support cone, 6 identical jack screw assemblies with gearboxes, drives 
and measuring systems, 12 u-joints in total and a CFRP platform. 
More technical details about these components are given in appendix \ref{appendix_components}.
Observations are started with extended jacks from a neutral position, Figure \ref{fig2} Right Panel.
The free access to receivers and correlator is shown in Figure \ref{fig4}.

\section{Correction Scheme and Control System: an  Overview}   \label{kinematics}

The hexapod topology is optimized with respect to minimized travel ranges of the u-joints and 
to minimized jack screw loads. 
Since the hexapod position is entirely determined by the variable length of the 6 jack screws together with 
the positions of the 6 lower fixed u-joints, 
utmost care needs to be taken to accurately monitor the jack screw lengths. The positions of the 6 lower
fixed u-joints have been measured with a laser ranging system in the Vertex factory in Germany.
The first group of pointing corrections on jack level
consist of 4 compensations: jack screw pitch error, temperature
compensation, jack screw rotation error and support cone correction.
They all directly yield a length correction for each individual jack at any given mount position.
These error compensations have been tested and calibrated in the Vertex factory, and the correction algorithms
are integrated into the hexapod control system.
Besides the group of jack screw corrections, a second group of telescope pointing corrections is implemented: radio and optical
refraction,
an optical telescope (OT) collimation error correction and an interpolation table (IT) for residual errors.
Only the latter two ones need to be measured, updated and handled by the operator. 
This second group leads to corrections in azimuth and elevation, which 
are then translated into jack corrections.
A detailed description of the pointing error model is in appendix \ref{appendix_pointing}.

For radio observations the OT collimation error correction is deactivated in the pointing error model.
This assumes that the offsets derived from the optical pointing (with OT collimation correction, optical
refraction and all other corrections activated) are identical to the errors for the radio pointing (no collimation 
error correction, radio refraction and all other corrections activated). In fact, the resolution and the collecting total power
of the 0.6m diameter antennas are not sufficient to do a separate radio pointing. However, we use the correlated signal
to verify the relative radio alignment 
%\newpage
%\noindent
between individual antennas.

We remark that we do not derive explicit corrections for the platform polarization
because the polarization stability and 
precision have been found to be around $0.1^{\circ}$ or better 
which is good enough for our purposes.
We, however, derive azimuth and elevation pointing corrections for various platform polarizations
 (section \ref{pointing_pol}).

The main drive control and the jack length calculations for a commanded position are done 
by the ACU (Antenna Control Unit). This also includes the inverse backward transformation giving
the telescope position based on a set of jack lengths. This is essential for the continuous check
between requested and actual telescope position, which is done in a closed loop system every 5ms 
and updated depending on the operation. The above mentioned pointing corrections 
are calculated on the PTC (Pointing Computer) where they 
can be individually activated and displayed. From here they are transferred to the ACU,
illustrated in the flow chart in Figure \ref{fig5}.

The block diagram in Figure \ref{fig6} summarizes the control system.
The actual position, defined as the real position 
after applying all the pointing corrections, is 
displayed on ACU and reported to the remote  TCS (Telescope Control System).
A redundant independent safety level is provided by the HPC (Hexapod Computer) with its PLC
(Programmable Logic Controller). 
The HPC calculates the telescope 
position from the jack positions as measured by the safety (auxiliary) encoders.
The PLC is responsible for safety interlocks from limit switches.
Time synchronization is derived from a stratum-1 GPS server,
connected to the ACU
with an IRIG-B time signal. Communication between individual computers 
is through standard LAN ethernet cables with TCP/IP protocol and NTP for time synchronization, 
and it is RS232 and/or CANbus
where analog components are involved. 

The telescope main operating modes include: {\it Preset, Startrack} and {\it Progtrack}, with the latter only possible 
in remote mode from TCS. {\it Preset} moves the telescope to a defined position in $(Az,El,hexpol)$ on a 
geodesic path, ensuring a short and fast connection between subsequent mount positions. {\it Startrack} 
tracks a celestial object with either $hexpol$=constant or $skypol$=constant. {\it Progtrack}
drives the telescope on a defined trajectory in $(Az,El,hexpol,time)$ with a spline interpolation and 
a maximum stack of 2000 data points. This mode is extensively used for various types of observation
and system checks.

\section{Performance Verification} \label{performance}

In the initial AMiBA operational phase seven close-packed 0.6m diameter Cassegrain antennas \citep{koch06} are used
on baselines separated by 0.6m, 1.04m and 1.2m. 
The antenna field of view (FWHP $\sim 23^{\prime}$) and the synthesized beam ($\sim 6^{\prime}$)
of the array in this configuration (at the observing frequency band 
86-102~GHz) set the specifications on the platform deformation and the 
pointing and tracking accuracy, which are: $\sim 0.6^{\prime}$ rms pointing error and a platform 
z-direction deformation of less than 0.3mm.

\subsection{Platform photogrammetry}  \label{photogrammetry}

Prior to the integration of the platform and the hexapod in Germany for the factory acceptance test, we performed 
stiffness measurements of the CFRP platform on the ground
under expected loading conditions. These measurements were repeated at the Mauna Loa 
site. Both measurements showed that the deformations were larger than expected and predicted by FEA 
(Finite Element Analysis), especially towards the outer edge of the platform,
even after reinforcement was added.
The cause is the segmented structure of the platform, coupled with inaccurate modeling of
stiffness across the segment joints.
It was decided to use the photogrammetry method to verify the platform deformations in a 
real 7-element compact configuration on the hexapod.

The first photogrammetry campaign took place during the fall of 2005, with dummy weights 
to replace receivers and electronic boxes. In October 2006, we repeated the photogrammetry measurements, 
this time on the operational 7-element telescope. The results of the second survey are consistent with the 
2005 results. In 2006, we achieved a better measuring accuracy: about 30$\mu$m rms in $z$ 
(defined as normal  
to the platform), with a short term (1-2 days) and a long term (1 week) repeatability better than 80$\mu$m rms.
We used a Geodetic Services, Inc. (GSI of Melbourne, Florida) INCA2 single digital photogrammetric camera. 
The pictures were processed with a GSI V-STARS 3D industrial measuring system. About 500 retro-reflective, 
self-adhesive targets (12mm diameter, $\sim$ 0.1mm thick)
 were distributed over the entire platform surface with a higher target density around 
receivers. 50 platform positions over the entire azimuth, elevation and platform polarization range
were surveyed. 

The analysis of the photogrammetry measurements \citep{raffin06,huang08b}
reveals a 
saddle-shaped platform deformation pattern at all surveyed positions, illustrated in 
Figure \ref{fig11}.
The amplitude and phase of this saddle are functions of azimuth, elevation and polarization. 
The specifications are met for the 7 antennas in the compact configuration, with 
a $z$-deformation amplitude up to 0.120~mm
in the inner part of the platform. 
At the location of the optical telescope (OT), 
the maximum amplitude (measured at 30$^{\circ}$ elevation and 20$^{\circ}$ polarization)
is about 0.38mm (Figure \ref{fig12}), which leads to an OT tilt movement with respect to the normal 
pointing axis of the mount
of about $\pm 1^{\prime}$ in this extreme position. A more average position $(Az,El,hexpol)$=(0,60,0)
is also illustrated for comparison, showing an amplitude of about 0.11mm, which leads to a corresponding 
OT tilt of about $0.25^{\prime}$. As argued in section \ref{pointing}, this uncompensated 
pointing error is acceptable
for the 7-element compact configuration, but will need to be corrected for the planned expansion phase with 
13 elements.

A detailed analysis and model of the saddle type deformation 
for a 13-element radio phase correction and an error separation 
between deformation and pointing error is presented in \citet{koch08a}.
For this refined analysis we installed a second OT on the platform. By simply taking the difference 
between the two data sets of the two OTs, a characteristic signature appears which can be clearly
attributed to the platform deformation. Using a saddle type model as an input for the deformation, 
the mount pointing error can be successfully separated. Similarly, with the help of an interpolation
scheme based on the entire photogrammetry data set, the radio phase error from the platform deformation
can be reduced from a maximum $800 \ \mu$m rms over the entire platform 
to $100\mu$m or less. The synthesized beam area
is then maintained to within 10\% of its non-deforming ideal value. 
In this way, the specifications are also met for the 13-element expansion.

\subsection{Hexapod optical pointing}   \label{pointing}

Pointing with a hexapod telescope is different from a conventional telescope where a 7- or 13-parameter pointing
error model is often used. 
An important consequence of the hexapod mount 
is the absence of azimuth and elevation encoders as compared to more 
traditional telescopes.  The 3-dimensional locations of the upper and lower 
u-joints in the measured reference 
positions and the jack lengths in any  position completely define the geometry of the mount. 
The pointing error model therefore needs to take utmost care 
in treating the jack lengths and the lower u-joint positions.
The 6 upper u-joint locations define a best-fit plane with its normal defining the resulting pointing axis.

Optical pointing is carried out with a Celestron C8 telescope, equipped with a Fastar f/1.95 adapter lens and a SBIG
ST-237 CCD camera. The resulting field of view (FoV) is about $30^{\prime}\times 20^{\prime}$ on a 652 $\times$ 495
pixel array, giving a calibrated pixel scale of about $3.81^{\prime\prime}$.
A preliminary rough OT collimation error is 
measured on the platform with a digital tiltmeter and 
compensated in the pointing error model, following equation~(\ref{collimation1}) and (\ref{collimation2}).

\subsubsection{Pointing with hexpol=0}

In order to achieve the required pointing accuracy of $0.6^{\prime}$ or better,  a two-step approach is adopted.
In a first pointing run all the known pointing corrections (section \ref{kinematics})
except the interpolation table (IT)
are activated on the pointing computer (PTC). 
As a function of the mount position, $(Az,El,hexpol=0)$, the offsets $(\Delta x_k$, $\Delta y_k)$ 
of a target star $k$,
with respect to the center of the CCD image, are split
into an $Az$ and $El$ error in a common reference frame at $Az=0$. This involves rotating the CCD images by the 
mount $Az$ (and $hexpol$) coordinate, together with an additional rotation $\gamma$ for the orientation of the CCD
 with respect
to the sky. This yields the CCD frame - mount frame transformation (Figure \ref{fig7}):
\begin{equation}
\left(\begin{array}{c} \Delta Az_{raw,k}\\ \Delta El_{raw,k} \end{array}  \right)= 
\left(\begin{array}{cc}  \cos\beta_k & \sin\beta_k \\
-\sin\beta_k & \cos\beta_k \end{array}\right)      \left(\begin{array}{c} \Delta x_k\\ \Delta y_k \end{array}  \right), 
\label{ot_equation}
\end{equation}
where $\beta_k=Az_k+hexpol_k+\gamma$ for each star image $k$ at the mount position $k$.
The raw errors $(\Delta Az_{raw,k},\Delta El_{raw,k})$ are then analyzed to separate the remaining 
uncompensated OT collimation error 
from the real mount pointing error.
 $(\Delta Az_{coll,OT},\Delta El_{coll,OT})$ has a characteristic azimuth and elevation signature of the form: 
\begin{equation}
\left(\begin{array}{c} \Delta Az_{coll,OT}\\ \Delta El_{coll,OT} \end{array}  \right)= \label{eq_ot} 
\left(\begin{array}{c} C_{Az}\\ C_{El} \end{array}  \right)+A   
\left(\begin{array}{c} \cos(Az+\phi)\\ \cos(Az+\phi+\pi/2)     \end{array}  \right),
\end{equation}
where $A$ and $\phi$ are the OT uncompensated tilt amplitude
and phase, respectively, 
which reflect the remaining OT collimation error and $(C_{Az},C_{El})$ are two constants.
Assuming a rigid OT, the amplitude $A$ is identical for the azimuth and elevation signature and 
their phases are separated  by $\pi/2$. 
The small fitting residuals $\Delta \bar{Az}_{IT}$, $\Delta \bar{El}_{IT}$ 
(of the order of $1^{\prime}$ or less) 
populate the IT, which is a three-dimensional table in $Az$, $El$ and $hexpol$:
\begin{eqnarray}
  \left(\begin{array}{c} \Delta \bar{Az}_{IT,k}\\ \Delta \bar{El}_{IT,k} \end{array}  \right)&=&\\ \nonumber
  \left(\begin{array}{c} \Delta Az_{raw,k}\\ \Delta El_{raw,k}           \end{array}  \right)&-&
\left(\begin{array}{c} \Delta Az_{coll,OT}\\ \Delta El_{coll,OT}       \end{array}  \right)+
\left(\begin{array}{c} C_{Az}\\ C_{El}                                           \end{array}  \right).
\end{eqnarray}

%\begin{equation}
%  \left(\begin{array}{c} \Delta \bar{Az}_{IT,k}\\ \Delta \bar{El}_{IT,k} \end{array}  \right)=
%  \left(\begin{array}{c} \Delta Az_{raw,k}\\ \Delta El_{raw,k}           \end{array}  \right)-
%\left(\begin{array}{c} \Delta Az_{coll,OT}\\ \Delta El_{coll,OT}       \end{array}  \right)+
%\left(\begin{array}{c} C_{Az}\\ C_{El}                                           \end{array}  \right).
%\end{equation}

The irregular grid errors ($\Delta \bar{Az}_{IT,k},\Delta \bar{El}_{IT,k}$) are then transformed into a 
regular spaced grid 
with the cubic Shepard algorithm \citep{renka99}, 
finally generating the IT pointing corrections ($\Delta Az_{IT},\Delta El_{IT}$).

In a second pointing run  all the known pointing corrections and the IT are activated on the PTC.
This verifies that the mount errors are reduced  with the IT and that only the OT collimation 
error remains.
 These small errors are checked every few weeks for their repeatability. 
Typically, one IT iteration is needed to reduce an initial pointing error in $Az$ and $El$ from $\sim 1^{\prime}$ rms to 
about $\sim 0.4^{\prime}$ rms. 
Subsequent pointing tests have revealed almost exactly the same numbers (within repeatability, section \ref{repeatability}),
so that our IT has been unchanged over the year of data-taking.
The iterative improvement with the IT is illustrated in Figure \ref{fig13}, 
where
the total raw error, $\sqrt{\Delta Az_{raw,k}^2+ \Delta El_{raw,k}^2}$, is shown in a polar plot over the entire azimuth 
range. 
In this strategy it is crucial to identify properly the remaining uncompensated 
OT collimation error ($\Delta Az_{coll,OT},\Delta El_{coll,OT}$) 
 to make sure that the small remaining values in 
the interpolation table compensate only and exclusively for the hexapod mount errors. 
For radio observations the OT collimation error correction is deactivated in the pointing error model.
This assumes that the pointing errors derived from the optical pointing (with OT collimation correction, optical
refraction and all other corrections activated) are identical to the errors for the radio pointing (no collimation 
error correction, radio refraction and all other corrections activated). No separate radio pointing is done
because the 0.6m antennas (FWHM $\sim 23^{\prime}$) 
do not have enough gain to allow us to verify the required pointing accuracy.
In order to have the 
most rigid measure,
the OT is installed close to one of the upper u-joint positions.  
Possible local irregularities and position-dependent platform deformations which can affect the rigidity of the OT 
need to be filtered out 
if the pointing needs to be further improved \citep{koch08a}.

The interpolation table approach 
further assumes that the remaining mount errors are sufficiently 
smooth enough functions in between neighboring pointings, 
leading to  the question of the optimized pointing cell size.  
We find that 100 stars, approximately evenly
distributed in solid angle over the entire accessible sky, resolve
the pointing features reasonably well. Observing more stars does not significantly improve the pointing. 
Typically, we need about 1 hour to observe 100 stars in a fully automatic mode,
where the telescope is driven from high to low elevation on a spiraling trajectory\footnote{
In the initial phase, the mount and pointing performance were extensively tested by driving 
the telescope manually from a few randomly chosen
initial positions to the same target position. These tests indeed helped to identify flaws in the control
software. Subsequently, with the stable control algorithm, different trajectories were found to be equivalent.  
For the automatic mode a spiraling trajectory was then adopted because
this leads to the most efficient sky coverage with a minimized overhead 
in telescope drive time.
}
.   
A multiple of this time is required if different platform polarizations for each star
are included.

Although not necessary in the interpolation table approach, 
we found it useful to further analyze the residuals and identify their origins.
A more detailed fitting including the mount tilt:
\begin{eqnarray}
  \left(\begin{array}{c} \Delta Az_{raw,k}\\ \Delta El_{raw,k} \end{array}  \right)&=&\\ \nonumber
  \left(\begin{array}{c} \Delta Az_{coll,OT}\\ \Delta El_{coll,OT}           \end{array}  \right)&+&
B\left(\begin{array}{c} \cos(Az+\psi)\times\sin(El)\\ \cos(Az+\psi+\pi/2)     \end{array}  \right),
\end{eqnarray}
where $(\cos(Az+\psi)\times\sin(El),\cos(Az+\psi+\pi/2))$ is a 
term taking into account an additional mount tilt, improved the goodness of the fit 
only marginally, but still revealed a mount cone/foundation tilt of $\sim 0.2^{\prime}$ with respect
to zenith. Furthermore, our control software allows us to simulate a small rotation of the entire telescope. In this 
way we identified a slight misorientation of the cone with respect to north of $\sim 1^{\prime}$.
These effects contribute partly to the constants $(C_{Az},C_{El})$ in equation~(\ref{eq_ot}). 
Since both errors are small, we 
simply absorb them in the IT.

\subsubsection{Pointing with hexpol$\ne$0}   \label{pointing_pol}

Extracting the polarization corrections relies on the proper identification of the OT signature. This is best done
at $hexpol=0$, since at $hexpol\ne 0$ the polarization error and the OT signature become degenerate. 
The $hexpol$ movement is illustrated in Figure \ref{fig16}.
For the OT itself, 
the $hexpol\ne 0$ case is only an additional rotation, 
identical to an azimuth position with $az+hexpol$, as described 
in equation~(\ref{ot_equation}). We are thus extracting the OT signature at $hexpol=0$
 assuming it to be rigid enough, so that 
any position error with a polarized platform, as a function of ($Az,El,hexpol$), becomes:
\begin{eqnarray}
  \left(\begin{array}{c} \Delta \bar{Az}_{IT,k}\\ \Delta \bar{El}_{IT,k} \end{array}  \right)&=&\\ \nonumber
  \left(\begin{array}{c} \Delta Az_{raw,k}\\ \Delta El_{raw,k}           \end{array}  \right)&-&
\left(\begin{array}{c} \Delta Az_{coll,OT}\\ \Delta El_{coll,OT}       \end{array}  \right)_{hexpol_k}+
\left(\begin{array}{c} C_{Az}\\ C_{El}                                           \end{array}  \right),
\end{eqnarray} 
where $(\Delta Az_{raw,k},\Delta El_{raw,k})$ are again defined as in equation~(\ref{ot_equation}) with 
$hexpol\ne 0$ and $(\Delta Az_{coll,OT},\Delta El_{coll,OT})_{hexpol_k}$ are the OT signatures shifted by $hexpol_k$,
$Az_k \rightarrow Az_k+hexpol_k$ in equation(\ref{eq_ot}).
This is illustrated in Figure \ref{fig17} for the raw elevation errors, where polarization 
pointing was done with $hexpol$=$\pm20, \pm10, 0^{\circ}$ for 100 stars. Clearly seen is the OT signature 
with the $hexpol=0$ case shifted to the different $hexpol$ angles. 
The residual polarization errors increase linearly with the polarization angle, from $\sim0.8^{\prime}$
to $\sim3^{\prime}$ for $0^{\circ}$ to $20^{\circ}$, as shown in Figure \ref{fig18} for
the polarization dependent elevation error.
We remark that we do not explicitly correct for a polarization error. We found that the uncertainty in the 
polarization angle is within $0.1^{\circ}$ or better, which is negligible for the later pointing error 
analysis and our observations. 

We finally consider the influence of the platform deformation (section \ref{photogrammetry})
on the pointing error analysis. The local saddle-type deformation with a z-direction amplitude of 
typically $100 \mu$m 
at the OT radius will slightly change the OT's phase and amplitude as a function of the mount position, and 
therefore introduce a local position dependent error which is under-/overcompensated in the IT. 
However, a  $100 \mu$m amplitude leads to an estimated wiggling of the OT of about $0.25^{\prime}$.
For the 7-element compact configuration we consider that acceptable and we therefore have not further
extracted this component.

\subsection{Repeatability}  \label{repeatability}

A key parameter to ensure the reliability of the entire system is the pointing repeatability.
Factory tests have revealed short term repeatability errors between  $1^{\prime\prime}$ and $7^{\prime\prime}$ 
in $Az$ and $El$\footnote{
For these tests a laser source was installed on the platform. The mount was repeatedly driven to
the same target position after going back to an initial position. The slight shifts in the locations of the projected
laser beams on the factory wall were then used to characterize the repeatability at the target positions. 
}
, respectively.
Two astronomical tests with the OT were performed: First, in an overall repeatability test,
two runs with 250 stars (distributed on equal solid angles over the complete accessible celestial sphere) 
were executed during the same night, the 2nd run immediately after the 1st one, with  
almost identical atmospheric conditions. 
Secondly, in a star position test, aiming at the day-to-day repeatability, 
a set of  8 stars in different directions and elevations was observed on two subsequent days at the same time. 
CCD images were then taken by going back and forth between two stars of this set.  
Some stars could be observed at almost exactly the same time at 
almost identical mount positions.
Both tests show consistent results:
A cell-to-cell comparison from the overall repeatability test
gives an average $Az$ and $El$ error difference of $5.7^{\prime\prime}$ and $2.5^{\prime\prime}$, respectively.   
The star position tests show a day-to-day repeatability 
better than $4^{\prime\prime}$. 
 From the same test it could also be verified that small changes in the mount position introduce 
only small linear deviations in the pointing and tracking errors. 

The long term repeatability has been checked on a roughly weekly basis in the early mount testing phase.
Later, with each additional receiver element integrated into the array, pointing tests have been 
routinely carried out. All these results are consistent within the short term repeatability errors. 
Over the more than two years operation of the 7-element compact configuration, a very robust pointing
performance was verified, with no measurable changes due to temperature or other environmental effects, 
or additional weight on the platform.

%\vspace{0.2cm}
\subsection{Tracking}

Tracking tests were performed over short time periods, typically about 30 minutes. 
This ensures that tracking results are not or only minimally biased by any uncompensated pointing errors.
30-minute tests mean that pointing stars remain within a single IT cell.
The tracking tests were done in  both polarization modes,
$hexpol$=constant and $skypol$=constant.\footnote{$hexpol$=constant: 
A celestial object is tracked by keeping the intrinsic platform polarization constant.
In this mode, any possible complication from the additional polarization movement  is avoided, 
making it a clean measure of tracking performance. 
In order to separate the tracking and pointing errors, tracking is done only over a short time 
(small angular range), where the pointing error is supposed not to change 
significantly. 
Longer tracking tests require the use of the IT.\\
$skypol$=constant: A celestial object is tracked by fixing a polarization on the sky, introducing     
 				therefore a counter-rotation of the hexapod to compensate for the sky rotation.
This checks the tracking stability of the mount. Constant
    $skypol$ mode is essential for polarization measurement and
    control of baseline orientations. We test this mode by monitoring
    the orientation of a vector between two stars.
}
In the $hexpol$=constant mode, a linearly increasing tracking error was measured over 30 minutes, 
accumulating to $20^{\prime\prime}$ after 30 minutes in Figure \ref{fig28}.
 The $skypol$=constant mode shows a field rotation stability of about $3.6^{\prime}$ rms with respect 
to an initial polarization direction over 30~minutes, Figure \ref{fig29}.
 Since tracking is internally handled as a fast
sequence of pointing positions, the long-term 
tracking error (several hours or large ranges in $Az$, $El$ and $hexpol$) 
is controlled by all the pointing corrections including the IT. On short timescales 
(within interpolated errors in the IT)
the measured tracking error reflects a combination of partially uncompensated pointing errors and tracking-specific 
errors\footnote{
Tracking-specific errors result from the moving telescope, contrary to the pointing errors where the telescope 
accuracy is measured at a fixed position. However, on a short time-scale tracking errors are typically small, and 
they are therefore not further taken into account in the analysis.
}.
Our normal observations \citep{wu08} 
are carried out in a lead/trail-main field procedure over about 6~minutes.
The tracking error is therefore negligible.

%\vspace{3cm}
\subsection{Array Efficiency}

Based on the measured pointing accuracy we quantify the array efficiency. Besides this absolute mount pointing
error $(p)$, which misaligns each antenna by the same amount, we also take into account the measured platform 
deformation error $(d)$ and a radio misalignment error $(m)$ for each antenna pair (baseline). 
$d$ and $m$ define a relative error which measures the shift in the overlap of two antenna primary beams.
Additionally, they can also alter the absolute pointing error of each antenna, worsening or improving it
due to the local pseudo-random character of the errors. 
This resultant total pointing error defines the loss in the 
synthesized beam product \citep{koch08a}. Due to the complicated position dependence of all these errors, 
we use a Monte-Carlo simulation (10,000 realizations) to quantify the loss. We assume uniformly 
distributed errors: $p\in [-0.4,0.4]^{\prime}$,  $d\in [-1,1]^{\prime}$,  $m\in [-2,2]^{\prime}$.
The error interval for $d$ is a conservative estimate from the photogrammetry results, $m$ corresponds
to $\sim 10\%$ of the antenna FWHP, which is approximately the achievable mechanical alignment limit. 
The expected efficiency is about 93\%.
With $p\in [-1,1]^{\prime}$ the efficiency drops to about 90\%.

%\vspace{0.1cm}
\section{Conclusion}           \label{conclusion}
The hexapod design has some interesting advantages compared to a conventional mount. From an engineering  
point of view, it offers a relatively compact design for transportation and possible relocation, a simple cable
wrap and ease of access to the receivers from beneath the platform, which is important for daily 
maintenance and easy reconfiguration of the baselines.  
For its astronomical application, there is no zenith keyhole, polarization movements are controlled by the same 
algorithm actuating all six jack lengths and no additional mechanical polarization axis is needed as in the 
case of the CBI telescope \citep{padin02}.
Compromises have to be made, leading to a reduced elevation and polarization range of  
30$^{\circ}$ and $\pm 30^{\circ}$, respectively. Whereas the hexapod offers six degrees of freedom, for its
traditional astronomical application only three degrees are used.
%with the reduced ranges being the cost of the 
%unused degrees of freedom.
%\footnote{
The AMiBA hexapod is essentially driven along geodesics on a sphere, both for tracking and slewing. 
Tracking is approximated as a sequence of multiple segments of great circles. Direct and even shorter travel ranges 
for slewing, following simple translational movements on a straight line, have not been implemented for safety reasons. 
The hexapod however offers
additionally the possibility of small movements along the pointing directions if this should be of astronomical
interest.
% }

Pointing with a hexapod is very different from a conventional mount. Position errors are not accumulated
as in a serial kinematics system, but the control system is significantly more complex. Utmost care 
needs to be taken to determine all the jack lengths accurately. With one IT iteration the current pointing 
accuracy is  $\sim 0.4^{\prime}$ rms in azimuth and elevation, which meets the requirements for the 7-element 
array compact configuration baselines. An uncompensated pointing error from the platform deformation of the 
order $\sim 0.25^{\prime}$ still keeps it in an acceptable range, but needs to be further analyzed for the
planned expansion phase. Position errors with polarization movements without IT compensation are typically
linearly increasing with polarization as a result of a slight drift of the platform center during
the rotation, which is a consequence of the absence of any mechanical polarization axis. This is also compensated
with the IT. 
Laser ranging techniques - along the jack screws or mounted on the ground to determine the orientation of the 
platform - might be a possible choice to further improve the pointing accuracy of a hexapod.  \\

%% The reference list follows the main body and any appendices.
%% Use LaTeX's thebibliography environment to mark up your reference list.
%% Note \begin{thebibliography} is followed by an empty set of
%% curly braces.  If you forget this, LaTeX will generate the error
%% "Perhaps a missing \item?".
%%
%% thebibliography produces citations in the text using \bibitem-\cite
%% cross-referencing. Each reference is preceded by a
%% \bibitem command that defines in curly braces the KEY that corresponds
%% to the KEY in the \cite commands (see the first section above).
%% Make sure that you provide a unique KEY for every \bibitem or else the
%% paper will not LaTeX. The square brackets should contain
%% the citation text that LaTeX will insert in
%% place of the \cite commands.

%% We have used macros to produce journal name abbreviations.
%% AASTeX provides a number of these for the more frequently-cited journals.
%% See the Author Guide for a list of them.

%% Note that the style of the \bibitem labels (in []) is slightly
%% different from previous examples.  The natbib system solves a host
%% of citation expression problems, but it is necessary to clearly
%% delimit the year from the author name used in the citation.
%% See the natbib documentation for more details and options.

{\bf Acknowledgment}
We thank the anonymous referee for providing useful comments and suggestions.
We thank the Ministry of Education, the National
Science Council, and the Academia Sinica for their support of this
project.  We thank the Smithsonian Astrophysical Observatory for hosting
the AMiBA project staff at the SMA Hilo Base Facility.  We thank the
NOAA for locating the AMiBA project on their site on Mauna Loa.  We
thank the Hawaiian people for allowing astronomers to work on their
mountains in order to study the Universe.  
We thank all the members of the AMiBA team for their hard work.
Support from the STFC for MB and KL is also acknowledged.

\appendix

\section{Hexapod Main Components} \label{appendix_components}

\subsection{Support Cone}

The support cone provides stiffness and inertia for the drive system. 
It consists of 3 inner and 3 outer truncated cone steel segments.
Individual segments are connected with each other with butt-strap joints for the highest stiffness. Corrosion
protection is assured with a 3-layer paint system. The 
anchoring is leveled to about $0.1^{\circ}$.  
Due to environmental 
and cost concerns associated with excavation on Mauna Loa,
the cone is not embedded but is sitting on the concrete 
foundation. A future cone insulation should further improve its thermal behavior.
A finite element analysis (FEA) was done in order to optimize a low structural weight and minimize the pointing
error contribution. Simulated load cases included gravity, 10 m/s side wind, 10 m/s front wind, a temperature
gradient along the cone axis with $\Delta T$=1K and a temperature difference between the steel cone and 
the concrete foundation with  $\Delta T$=2K. In order to separate the pointing error contribution of
the support cone from the rest of the telescope, the entire structure above the cone was modeled
to be perfectly rigid with no reaction forces. The wind loads were treated as resultant nodal point forces
on the platform gravity center. 
The FEA demonstrated that the contact area between the support
cone and concrete foundation is everywhere under pressure in any
operational state. 
The main pointing error contribution then comes from the gravity 
load with maximum errors 
at lowest elevation and maximum polarization of about $8^{\prime\prime}$ and $3^{\prime\prime}$
 in azimuth and elevation, respectively. 
Temperature gradients (within the cone and between cone and concrete ground) contribute in total
about $1^{\prime\prime}$ to azimuth and elevation pointing errors.
10m/s front and side winds 
can give up to $1^{\prime\prime}$ pointing error contribution. Generally, the  
pointing error in both azimuth and elevation increases 
by about $4^{\prime\prime}$ to $8^{\prime\prime}$ if the polarization is changed from 
$0^{\circ}$ to $30^{\circ}$.
These errors are calibrated with the help of an interpolation table (appendix \ref{appendix_pointing}).

\subsection{Universal Joints and Jack Screws}

Very stiff and backlash-free u-joints
are necessary in order to meet the 
pointing requirement, because of the high torques under drive
conditions, the large shear forces at low elevation, and because
u-joint deformation cannot be detected directly. Zero backlash is
achieved by tapered roller bearings which are preloaded in the
axial direction.
A large angular range of motion (partly up to $\pm$ 52$^{\circ}$) in 
tangential and radial direction is necessary to achieve the telescope travel range. Minimizing the travel range
of the u-joints and jack screws by keeping small dimensions, low weight and high stiffness has been one of the 
key design achievements. The end positions of the u-joints are monitored by limit switches which shut off
the servo system. 

The jack screws consist of a tubular ball screw with an integrated low backlash worm gear with a transmission 
ratio 10.67:1. Each jack is driven by a motor and a low backlash bevel gear with transmission ratio 4:1.
The ball screw spindle of the jack screw is engaged in the worm
gear by a backlash-free axial preloaded double nut, which is also free from axial backlash. 
The jacks, each with a maximum operation load of 100 kN, 
 can be driven at a stroke rate of 0 to 20 mm/s.
An absolute (main)
angular encoder\footnote{
The encoder resolution is 13 bit $\times$ 4096 revolutions with a $\pm$1 bit accuracy, 
which leads to a $0.23\mu m$ overall resolution for a jack pitch of 20mm.
According to the manufacturer's specifications, the measurement inaccuracy is $< 0.1\mu m$,
which has a negligible influence on the pointing error.
}
is mounted at the shaft of the worm gear for an accurate measurement of the 
jack screw length. If necessary, this can be upgraded with 
a laser interferometer for a direct measurement of the actual jack screw
length, to compensate for errors in the 
jack screw pitch, the jack elastic deformation and the jack screw and worm gear backlash.
It is estimated that this would improve the pointing accuracy from $\sim10^{\prime\prime}$ to about 
$\sim3^{\prime\prime}$.
Since a hexapod design does not allow for hardware switches to limit the travel range of a telescope,
 a second set of encoders (auxiliary encoders) has been included for independent determination of jack and 
telescope positions in a separate safety controller, the HPC (Figure \ref{fig6}). 

\begin{figure*}
\begin{center}
\includegraphics[scale=1]{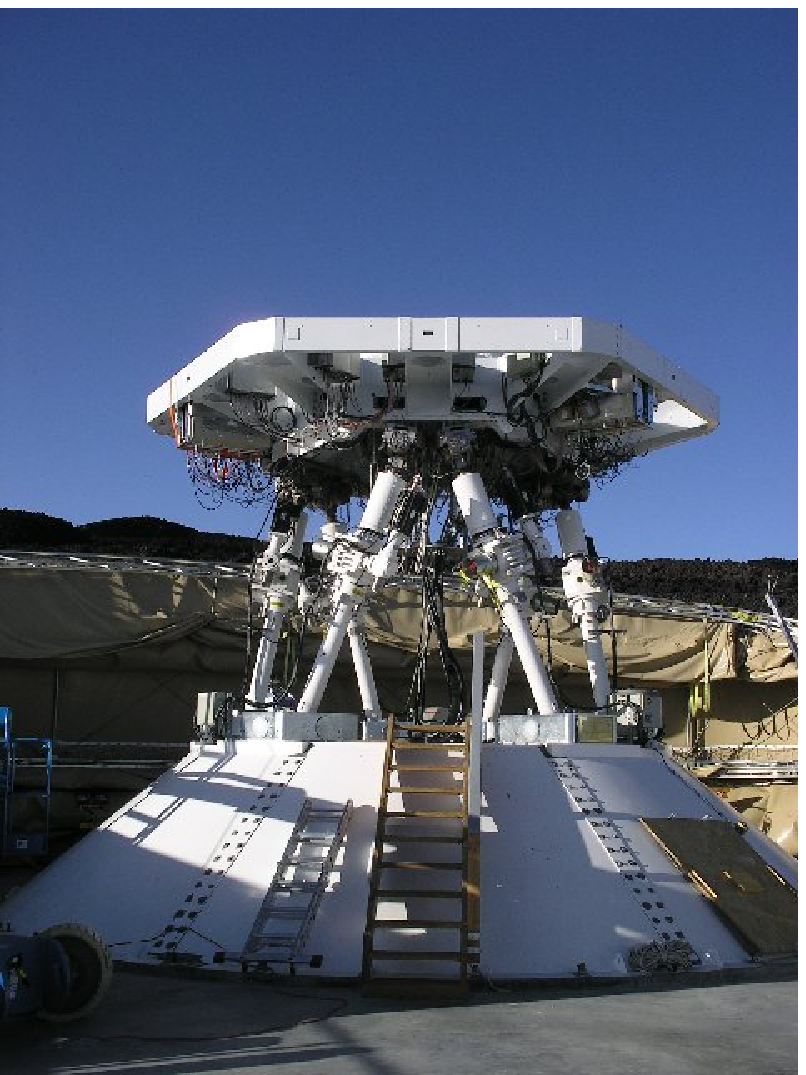}
\includegraphics[scale=1]{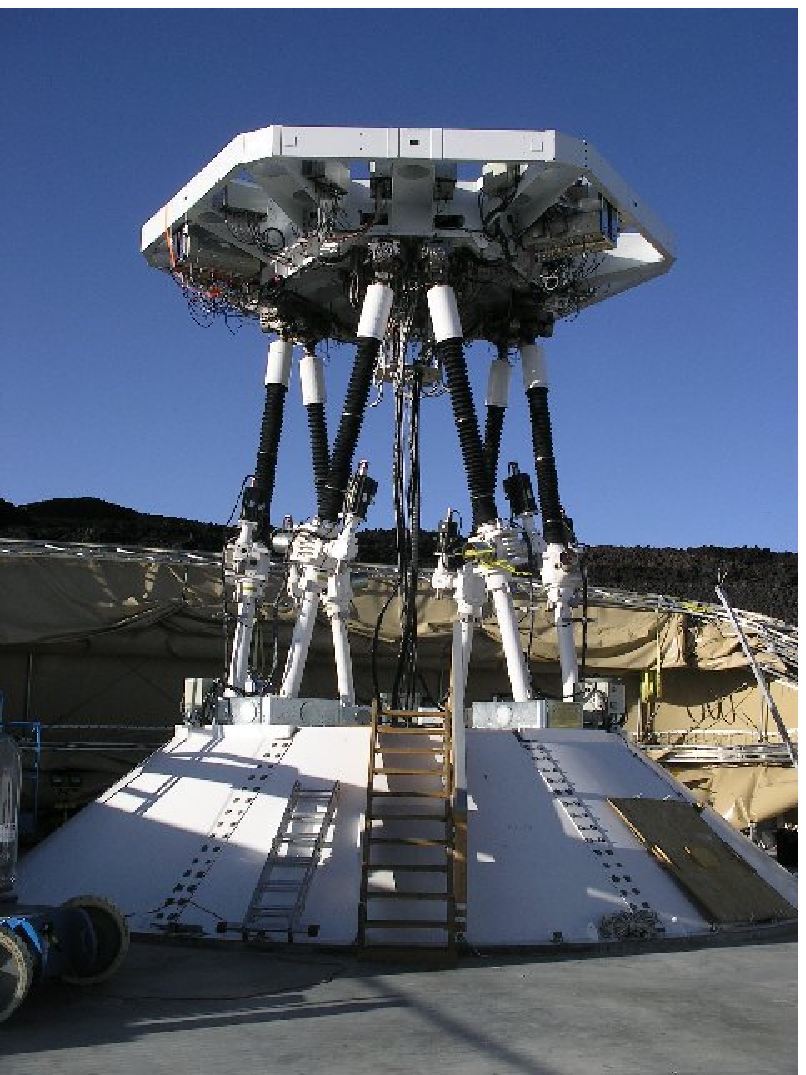}
\caption{\label{fig2} Left Panel: The AMiBA in stow position with fully retracted jack screws of about 2.8m length. 
In the back is the retractable
shelter to protect the telescope. The height of the hexapod is about 5.5m above ground level.
Right Panel: The AMiBA in neutral position with extended jacks to start the observation. 
The jack lengths are about 4.8m and the telescope height is about 7.5m above the ground.
}
\end{center}
\end{figure*}

\begin{figure}
\begin{center}
\includegraphics[scale=0.6]{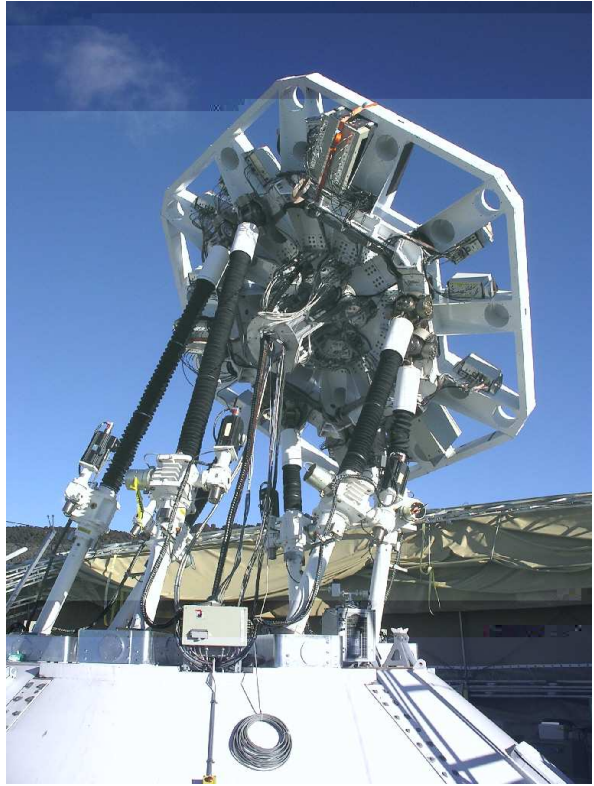}
\caption{\label{fig4}
Rear view of the AMiBA showing the free access to all the receivers. Cables and helium lines are guided
with a central fixed wrap in order to minimize the cable movement.
A correlator box (topmost) and various receiver electronic boxes are mounted on the outer spokes of the platform. }
\end{center}
\end{figure}

\begin{figure*}
\begin{center}
\includegraphics[scale=0.5]{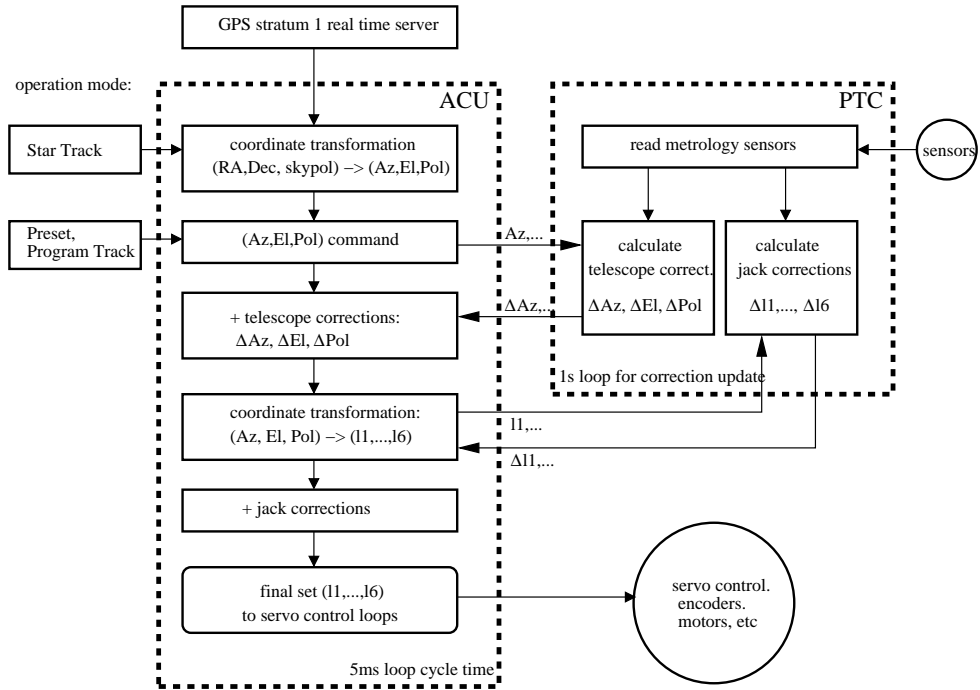}
\caption{\label{fig5}
The flow chart illustrating the data transfer for the pointing corrections between the 
Antenna Control Unit (ACU) and the Pointing Computer (PTC).}
\end{center}
\end{figure*}

\begin{figure*}
\begin{center}
\includegraphics[scale=0.5]{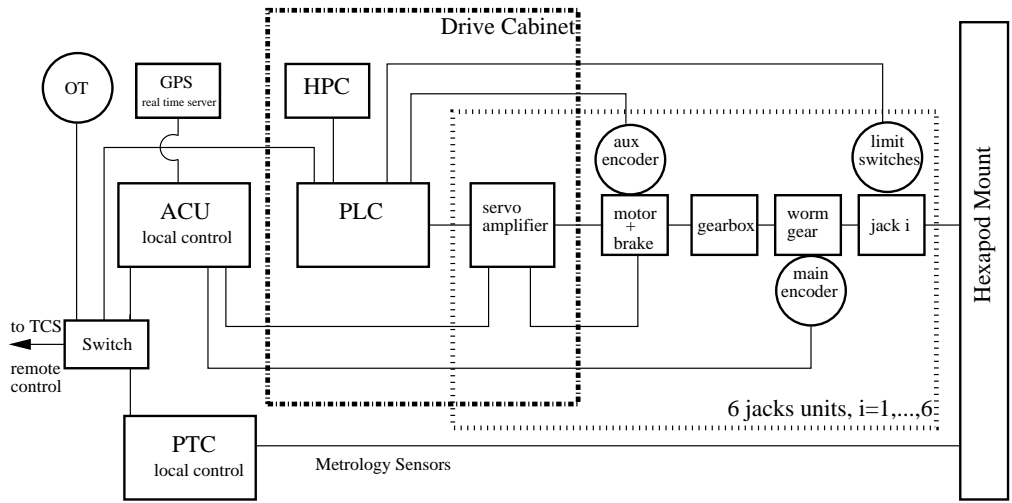}
\caption{\label{fig6}
Block diagram of the local control system.}
\end{center}
\end{figure*}

\begin{figure}
\begin{center}
\includegraphics[scale=0.55,angle=0]{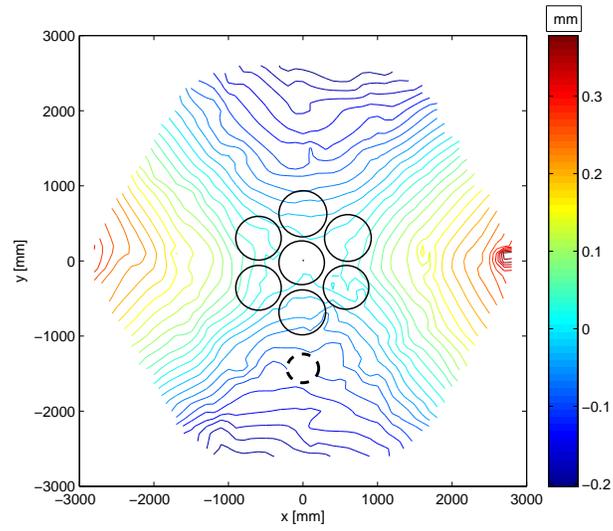}
\caption{\label{fig11}
Contour plot of a typical platform saddle deformation in $z$ direction at the mount position 
($Az,El,hexpol$)=(0,50,0), in units of mm.
The maximum deformations are  about +0.3 mm and -0.2 mm at the outer edges of the platform.
A roughly linear decrease in $z$ deformations is seen towards the center of the platform.
The locations of the OT (dashed circle) and the antennas (solid circles) are indicated for 
illustration. 
}
\end{center}
\end{figure}

\begin{figure}
\begin{center}
\includegraphics[scale=0.35,angle=-90]{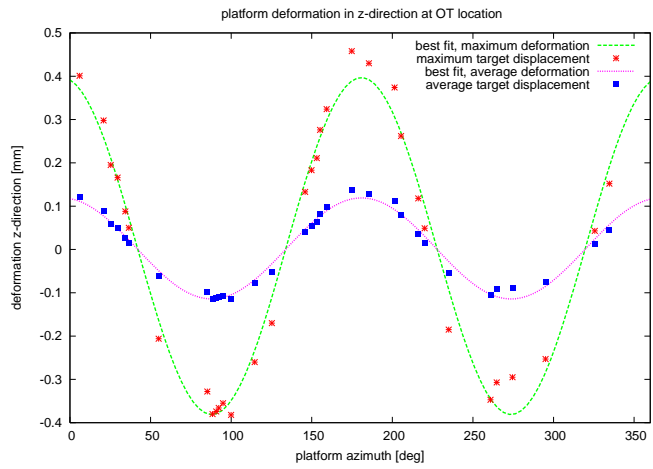}
\caption{\label{fig12}
Platform deformation in the $z$ direction as a function of platform azimuth 
at a radius r=1.4m where the OT is mounted. The selected data points are within an annulus
of r=$1.4m \pm 0.2$ {\rm m}. The maximum deformation
is measured at the extreme position ($Az,El,hexpol$)=(0,30,20) and sets therefore an upper limit to 
this uncompensated pointing
error. Also illustrated for comparison is an average mount position ($Az,El,hexpol$)=(0,60,0) with an 
amplitude of about 0.11mm, leading to a corresponding OT tilt of about $0.25^{\prime}$.   
Clearly visible is again the saddle structure with a functional form $\cos(2Az)$. 
}
\end{center}
\end{figure}

\begin{figure*}
\begin{center}
\includegraphics[scale=0.35,angle=0]{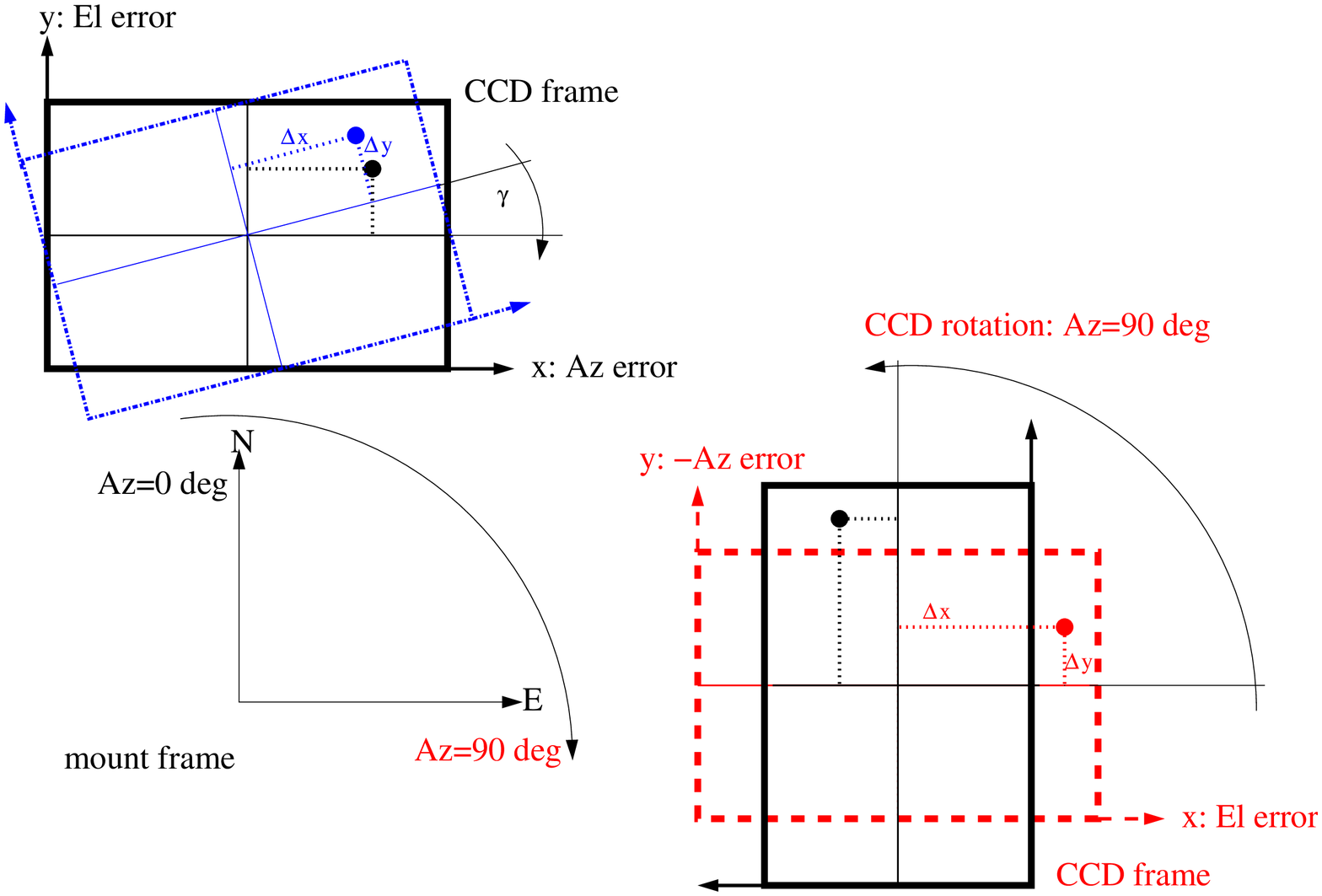}
\includegraphics[scale=0.35,angle=0]{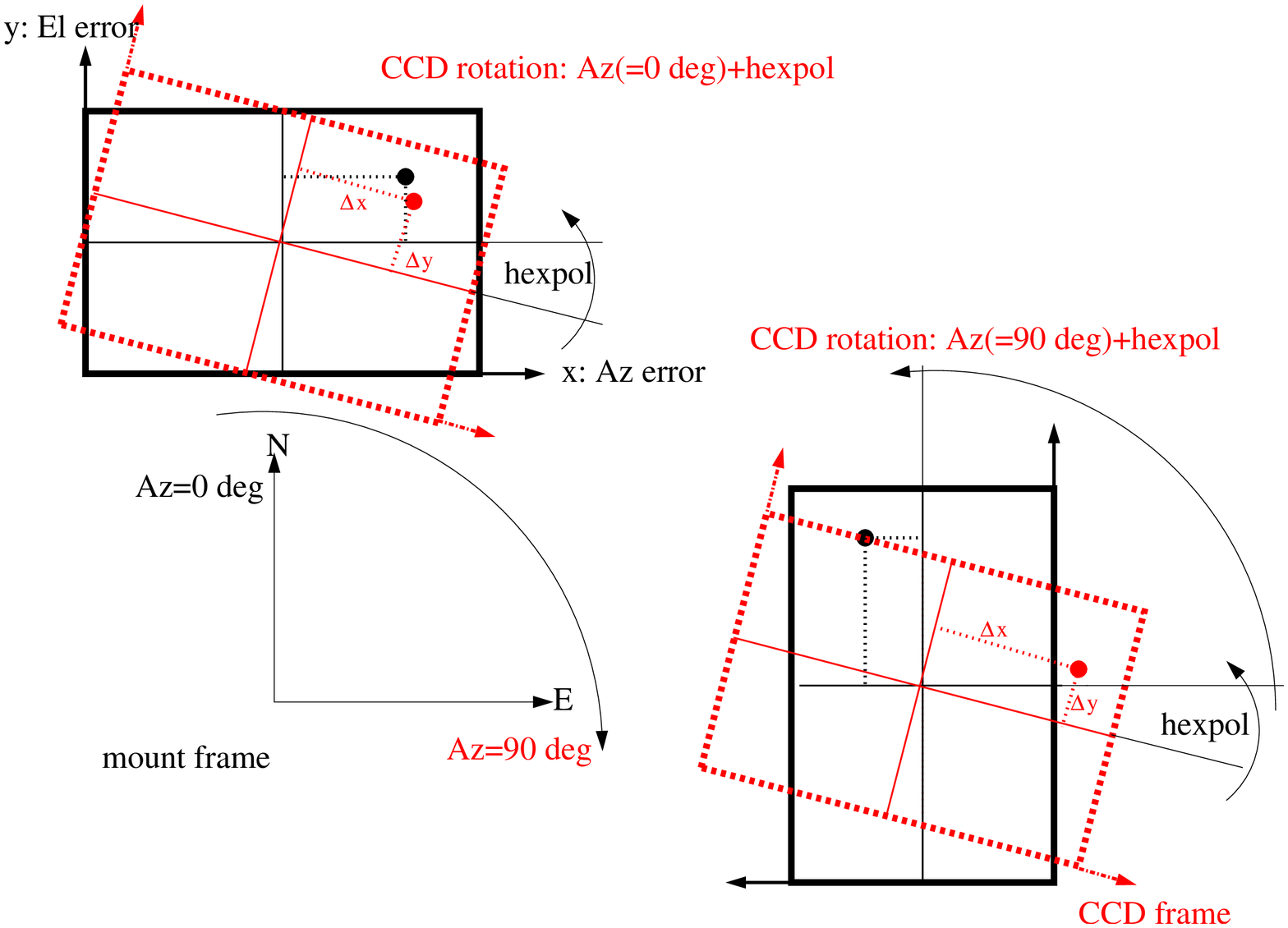}
\caption{\label{fig7}
Illustration of the CCD frame - mount frame transformation. 
Measured star position offsets are indicated with $\Delta x$ and $\Delta y$.
Left Panel: The rotation $\gamma$ aligns the CCD frame with the east-west 
sky drift direction (blue - black frames). At the reference position $Az=0$
the CCD axes then directly measure azimuth and elevation errors.
Moving to different azimuth positions only tips (but not rotates) the CCD frame, 
illustrated at $Az=90$ with the red dashed CCD frame. A rotation of $90^{\circ}$
brings it back to the reference position. The original blue frame before sky drift
alignment is omitted here.
Right Panel: Additional platform polarization ($hexpol$) which rotates the 
CCD frame (red dashed). A rotation of $Az+hexpol$ brings them back to the 
reference position. The blue frame is again omitted here for clarity.
}
\end{center}
\end{figure*}

\begin{figure*}
\begin{center}
\includegraphics[scale=0.7,angle=0]{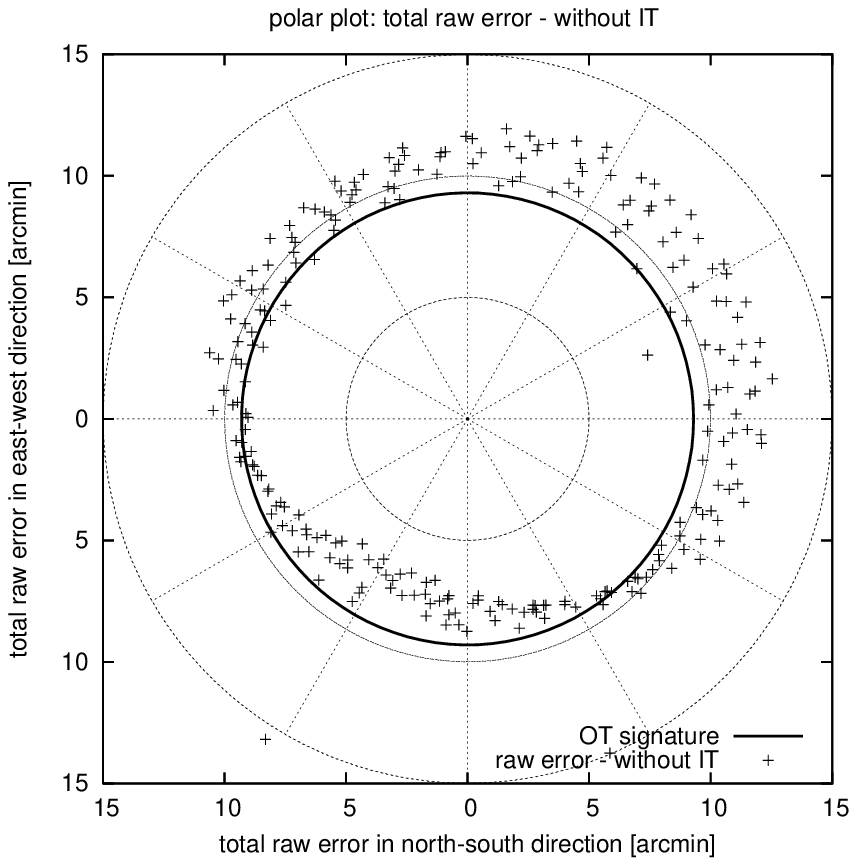}
\includegraphics[scale=0.7,angle=0]{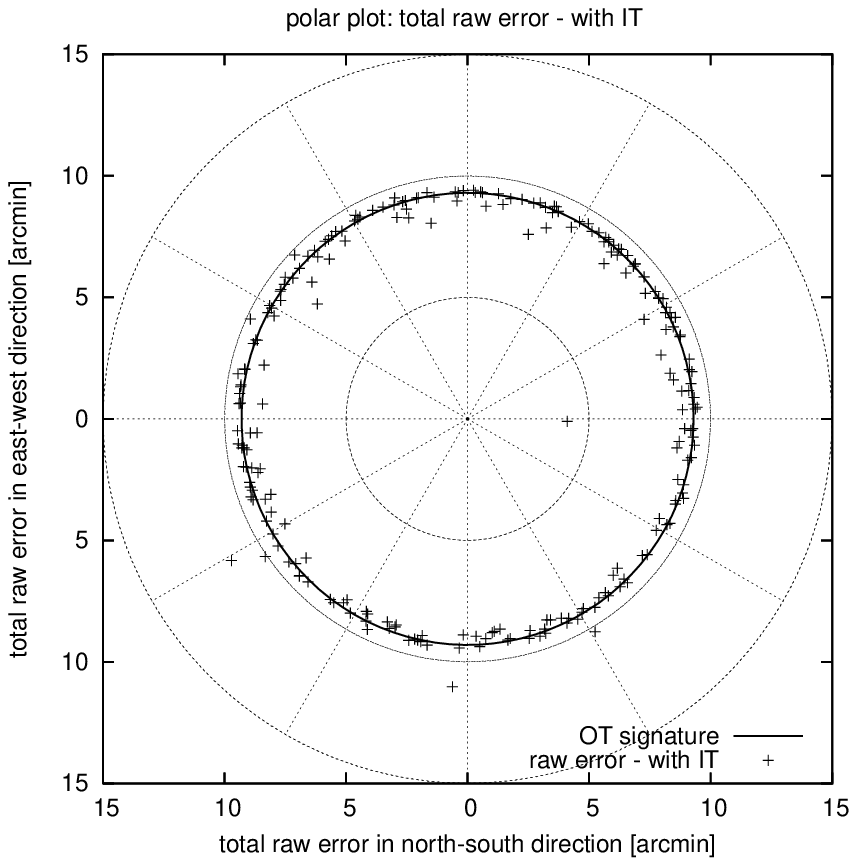}
\caption{\label{fig13}
Left Panel: Polar plot of the total raw error, combining $\Delta Az_{raw,k}$ and $\Delta El_{raw,k}$, 
without IT correction. 
The remaining uncompensated OT collimation error (OT signature) is shown as a 
circle which represents the OT tilt $\sim 9^{\prime}$ with respect to the mount pointing axis.
The data points scattered off the circle show the residual pointing errors $\Delta \bar{Az}_{IT,k}$
and $\Delta \bar{El}_{IT,k}$ for the IT, leading to $\sim 1^{\prime}$ rms pointing error. 
The test was done with 250 stars.
Right Panel: Polar plot of the total raw error with IT correction. 
The OT collimation error is again shown as a 
circle.
The reduced scatter compared to the left panel
verifies the improved pointing with $\sim 0.4^{\prime}$ rms pointing error.
}
\end{center}
\end{figure*}

\begin{figure*}
\begin{center}
\includegraphics[scale=1]{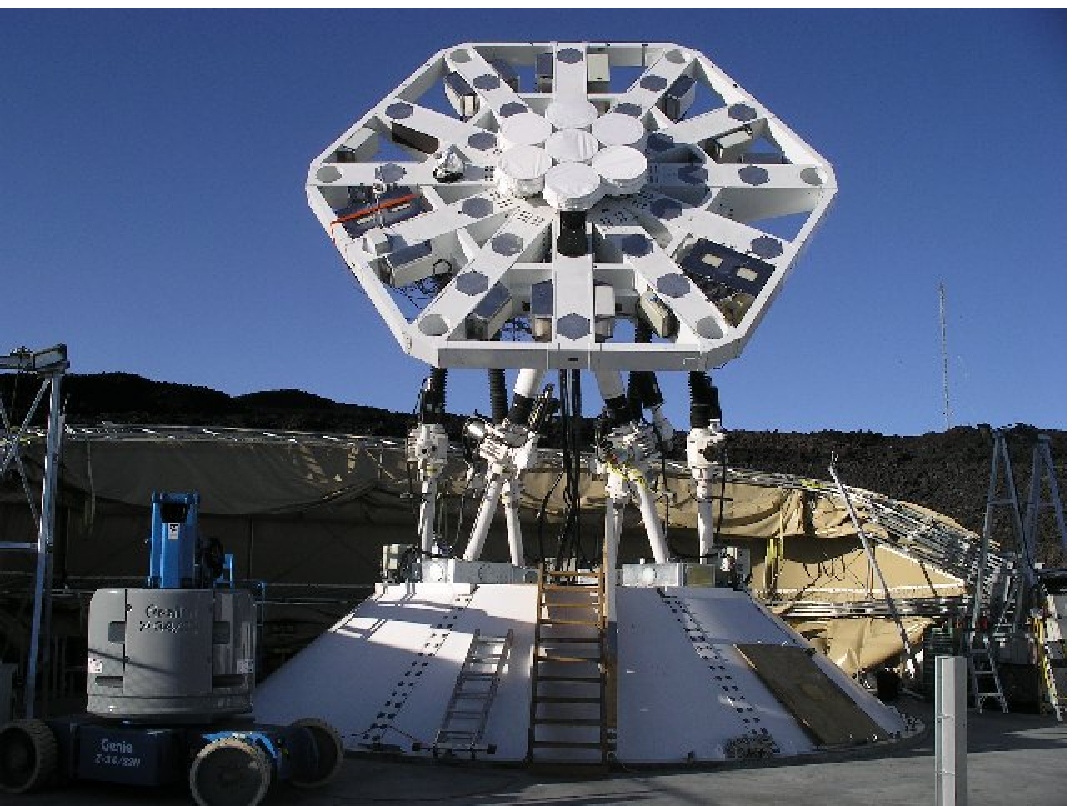}
\includegraphics[scale=1]{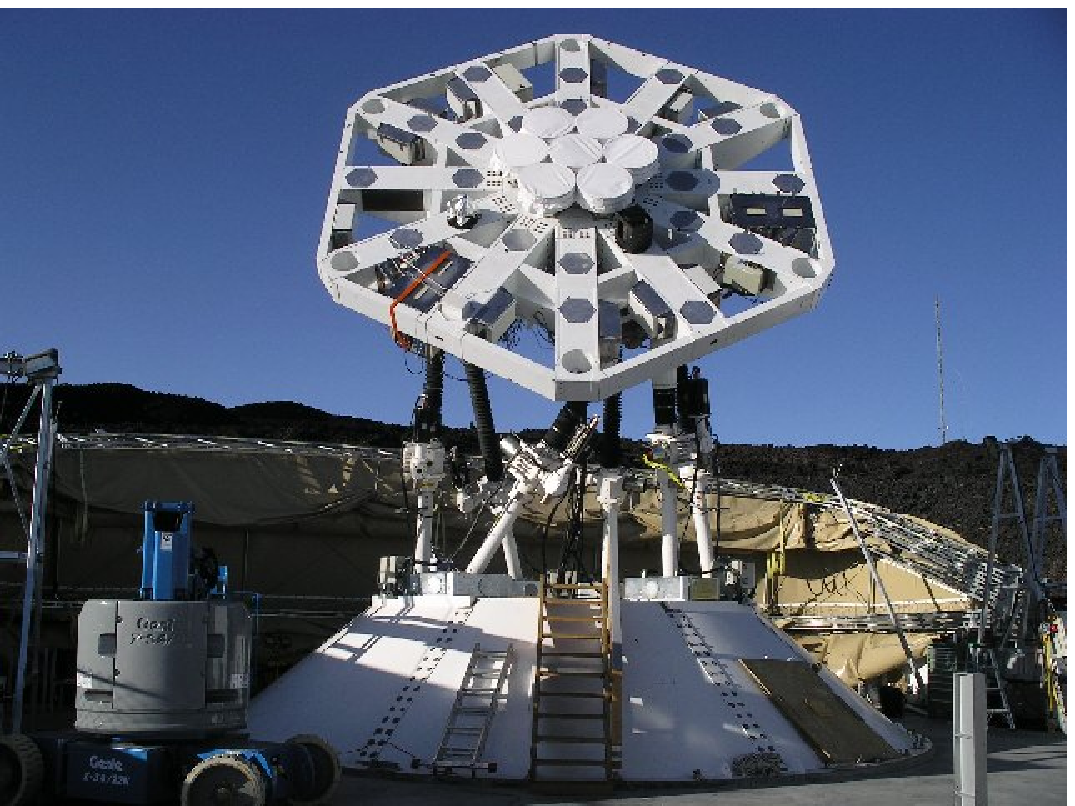}
\caption{\label{fig16} Left Panel: Front view of the AMiBA with a platform polarization $hexpol=0$.
Installed are 7 antennas in compact configuration, giving 0.6m, 1.04m and 1.2m baselines.
Free receiver holes in the CFRP platform allow for different array configurations and for a 
planned expansion phase with 13 antennas. The 8 inch refractor for optical pointing is attached to  
the black bracket below the lowermost antenna at a distance of 1.4m from the platform center.
Right Panel: Front view of the AMiBA with a platform polarization $hexpol=30$.
(Counter-clockwise rotation with respect to the left panel.)
}
\end{center}
\end{figure*}

\begin{figure}
\begin{center}
\includegraphics[scale=0.63,angle=0]{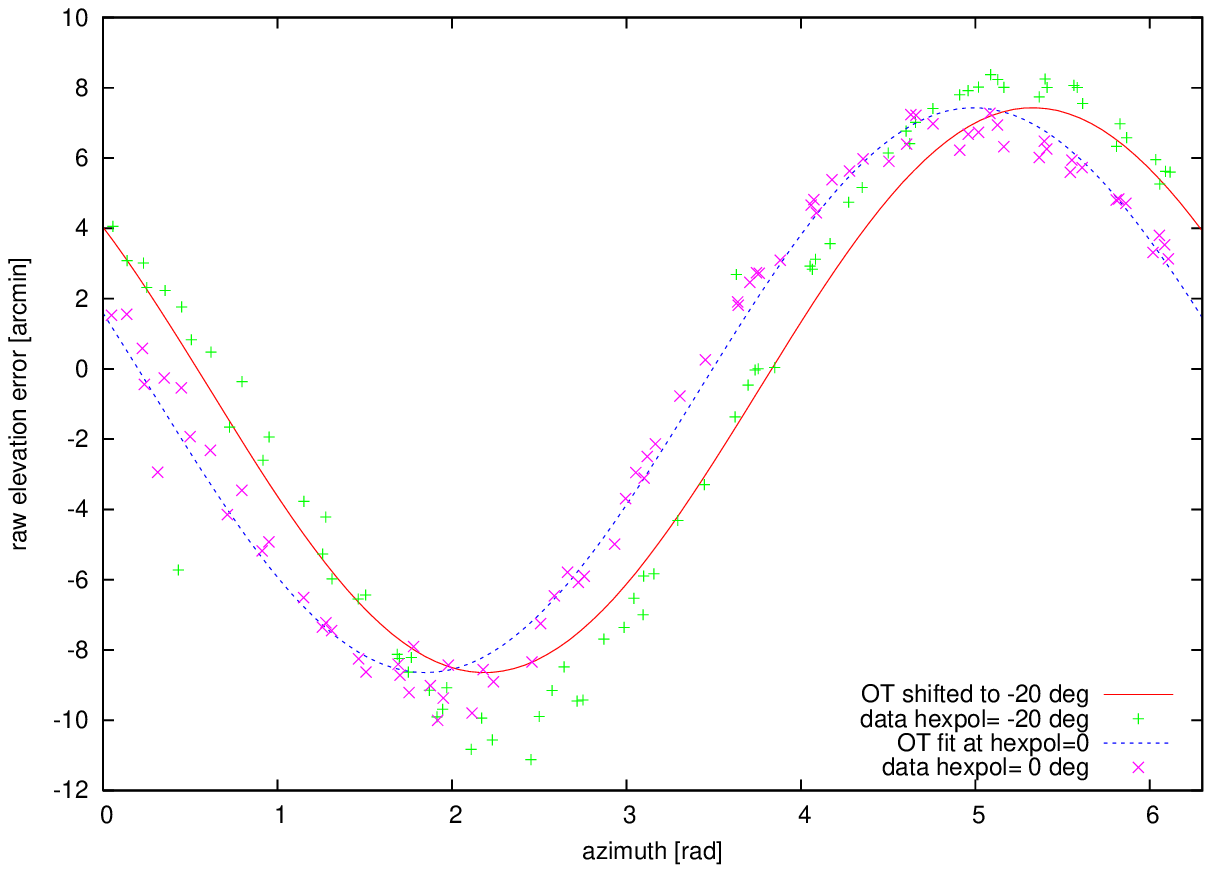}
\includegraphics[scale=0.63,angle=0]{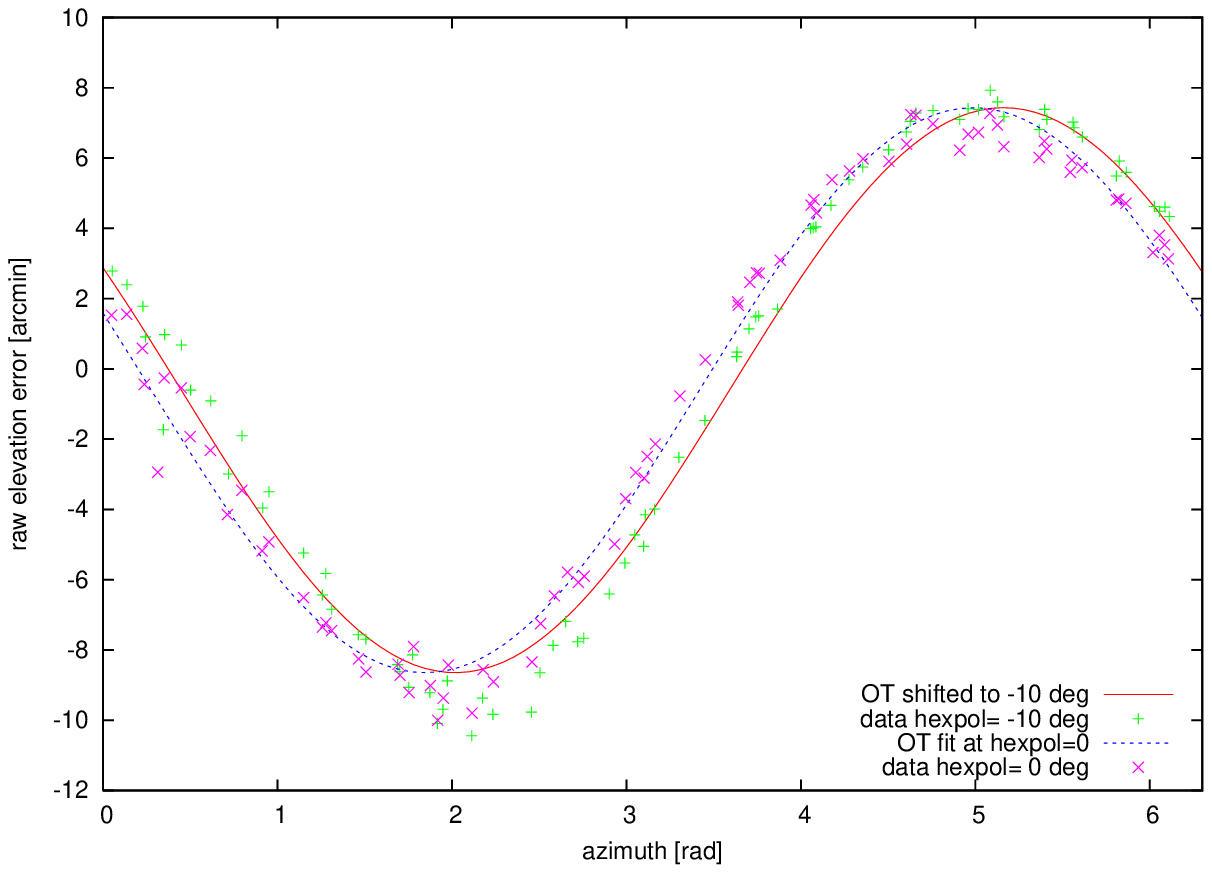}
\includegraphics[scale=0.63,angle=0]{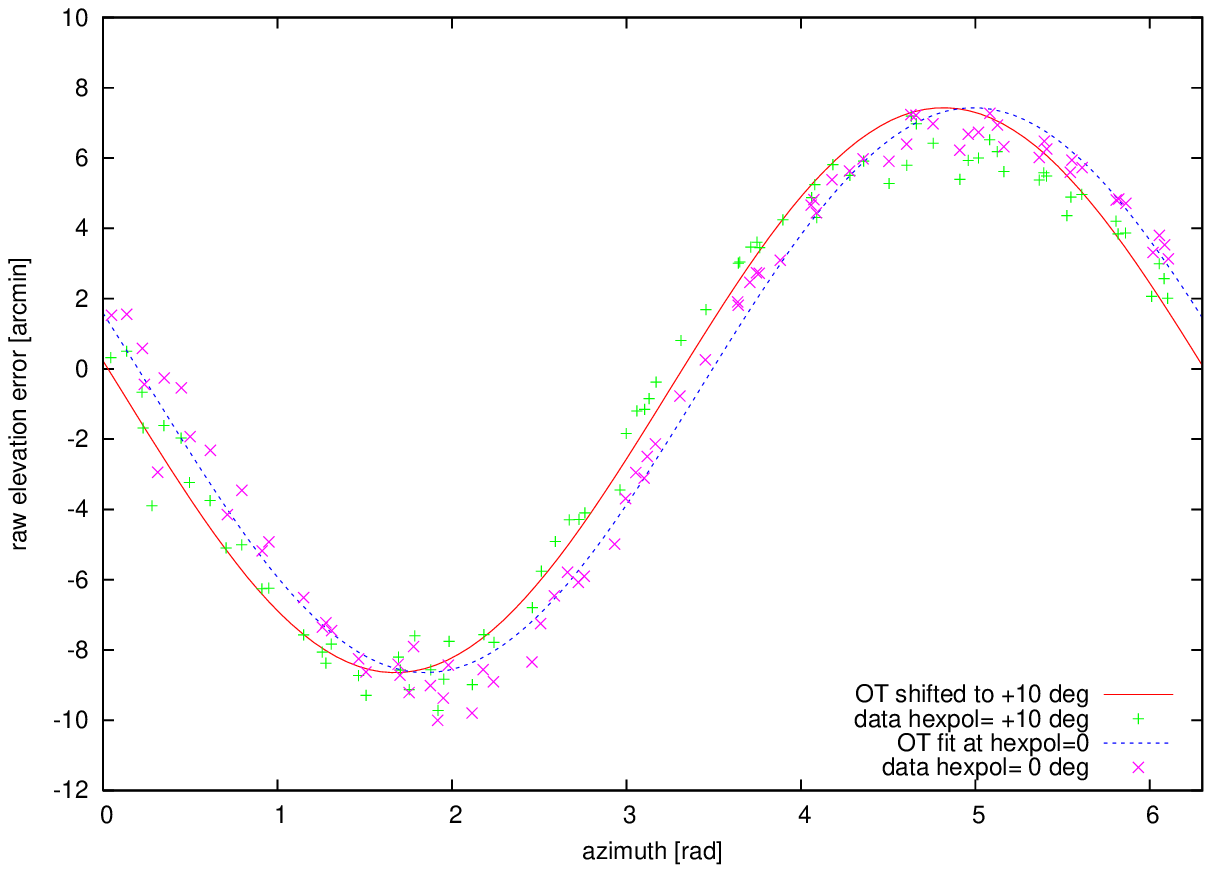}
\includegraphics[scale=0.63,angle=0]{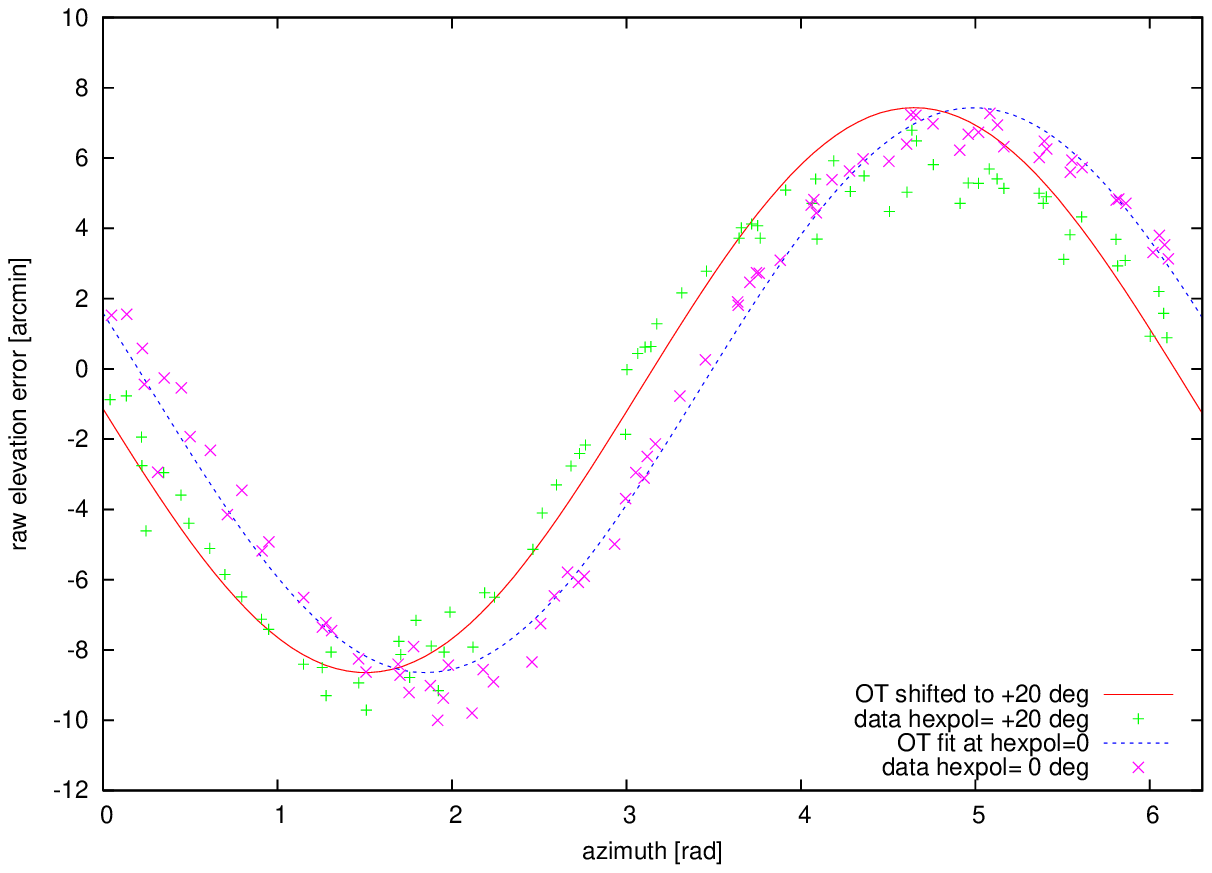}
\caption{\label{fig17}
Raw elevation errors (including remaining OT collimation error apparent as a cosine-like
signature) for $hexpol=-20, -10, +10, +20^{\circ}$, displayed in the panels 
from left to right and top to bottom.
An OT signature extracted from the $hexpol=0$ pointing data is shifted to the $hexpol\neq 0$ data to calculate
the residual errors for the IT.
}
\end{center}
\end{figure}

\begin{figure}
\begin{center}
\includegraphics[scale=0.35,angle=-90]{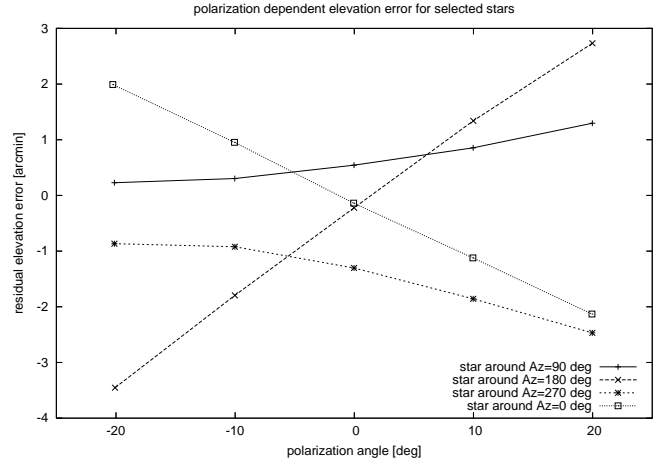}
\caption{\label{fig18}
Elevation error residuals after extracting the OT signature for selected stars at 
Az=$0,90,180,270^{\circ}$ as 
a function of $hexpol$. Errors tend to increase linearly towards larger polarization angles. The azimuth errors
(not shown) manifest the same result.}
\end{center}
\end{figure}

\begin{figure}
\begin{center}
\includegraphics[scale=0.35,angle=-90]{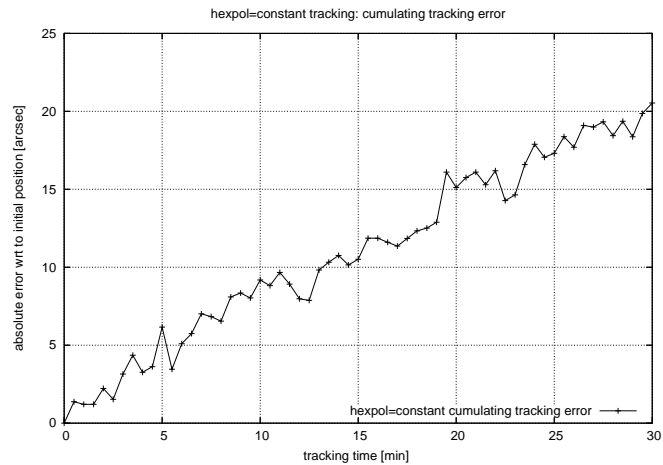}
\caption{\label{fig28}
Tracking accuracy in $hexpol$=constant mode, accumulating linearly to a tracking error of 
about $20^{\prime\prime}$ after 30 min.}
\end{center}
\end{figure}

\begin{figure}
\begin{center}
\includegraphics[scale=0.35,angle=-90]{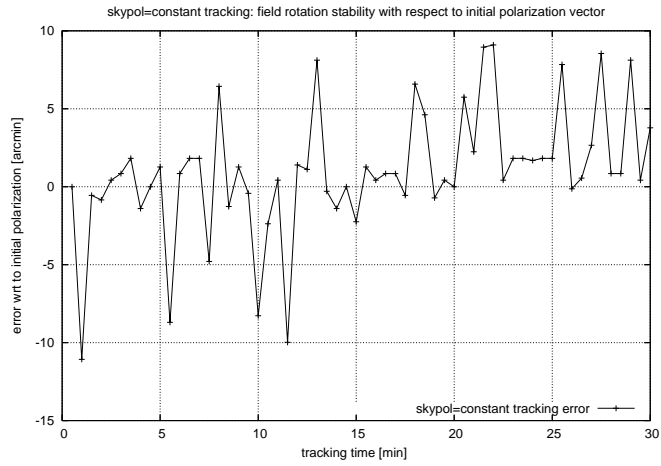}
\caption{\label{fig29}
Field rotation stability test in $skypol$=constant mode, showing about $4^{\prime}$ rms accuracy in 
the control of the 
rotation angle. Subsequent images are separated by 30 seconds.}
\end{center}
\end{figure}

\begin{figure*}
\begin{center}
\includegraphics[scale=0.4]{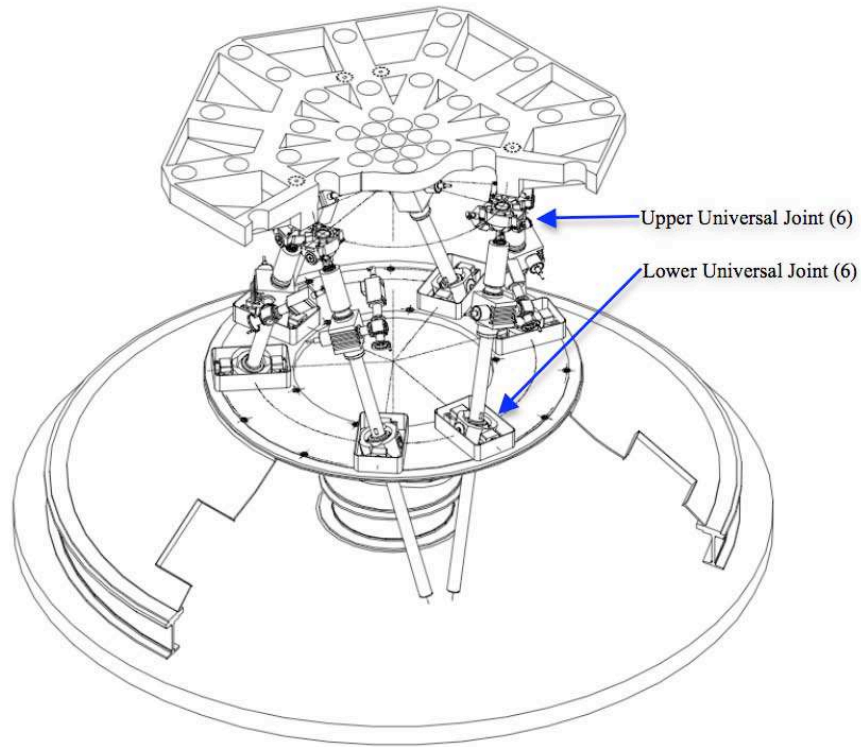}
\caption{\label{fig1} General design of the AMiBA hexapod mount with its main 
components: support cone, jack screws, u-joints and platform.}
\end{center}
\end{figure*}

The jack screws have a minimum and maximum length of about 2.8m and 6.2m, respectively, with a maximum travel 
range of 3.4m. 
They can be fully retracted to bring the telescope into a stow position (Figure \ref{fig2} Left Panel), which allows us to
close the shelter and protect the telescope and instruments 
from inclement weather and windy conditions. Observations are started with 
extended jacks from a neutral position, Figure \ref{fig2} Right Panel.

\subsection{CFRP Platform}

The AMiBA platform was designed by ASIAA and fabricated by CMA 
(Composite Mirror Applications, Inc.), Tucson, Arizona, USA. 
In a segmented approach, the central piece and the 6 outer elements are bolted together.
There are 43 antenna docking positions, allowing for multiple
configurations for the array, with baselines from 0.6 to
5.6~m. The receivers are sited behind the antennas and fit through
apertures in the platform.
The free access to receivers and 
correlator is shown in Figure \ref{fig4}. Cables and helium lines are guided with a 
central fixed cable wrap at the back of the platform. 
An optical telescope for optical pointing tests has been attached to one of the free 
receiver cells near the upper u-joints. Two photogrammetry surveys in 2005 and 2006 \citep{raffin06}
have revealed that the platform deformation during operation is within the specifications
of the 7-element compact configuration (section \ref{photogrammetry}).

\section{Pointing Error Model}   \label{appendix_pointing}

The pointing corrections on jack level consist of 4 compensations\footnote{Load cells were considered to measure the 
elastic jack screw deformation due to gravity. Estimates including a finite element analysis predict a maximum 
length error due to the bending of about $160 \ \mu$m only. The elastic deformation error is therefore not explicitly
modeled as a pointing error component, but simply absorbed in the error interpolation table.}
: {\it jack screw pitch error, temperature
compensation, jack screw rotation error}  and {\it support cone correction}.

A jack pitch error correction is essential since 
the real length of a jack is not directly measured. Only the rotation 
of the worm gear is measured.
Each jack has therefore been calibrated with a correlation function for the exact length against jack 
screw pitch error compared
to the encoder readout. The repeatable measurements are fitted with a polynomial of order 10, leading to a 
pitch error length correction $\Delta L_{p,i}$ for each jack $i=1,...,6$, coded in the software:

\begin{equation}
\Delta L_{p,i}\sim\sum_{k=1}^{10}a_{ik}\left (x_i\right)^k,
\end{equation}
where $a_{ik}$ are the fitting coefficients and $x_i$ is the jack length.

The jack screw length variations due to temperature changes are monitored with 3 temperature sensors along
each jack. The resulting change in length $\Delta L_{T,i}$ is calculated as:

\begin{equation}
\Delta L_{T,i}=\int_0^{l_i} \alpha f_i(x)\,dx,
\end{equation}
where $\alpha=12.0 \times 10^{-6}$K$^{-1}$ is the linear thermal expansion coefficient for ordinary steel 
and $l_i$ is the position
dependent length of the jack $i$. $f_i(x)$ is a linear approximation to the temperature distribution along the jack $i$:

\[ f_i(x)=\left\{ \begin{array}{r@{\quad:\quad}l}
\Delta T_{1,i} & x\le P_{1,i},\\
\frac{\Delta T_{2,i} - \Delta T_{1,i}}{P_{2,i}-P_{1,i}}(P_{2,i}-x) + 
\frac{\Delta T_{1,i}\, P_{2,i} - \Delta T_{2,i} \, P_{1,i}}{P_{2,i} - P_{1,i}} & P_{1,i} < x \le P_{2,i},\\
\frac{\Delta T_{3,i} - \Delta T_{2,i}}{P_{3,i}-P_{2.i}}(P_{3,i}-x) + \frac{\Delta T_{2,i}\, P_{3,i} - \Delta T_{3i} \, P_{2,i}}
{P_{3,i} - P_{2,i}} & P_{2,i} < x \le P_{3,i},\\
\Delta T_{3,i} & x > P_{3,i},
\end{array} \right. \]
where $P$ are the positions of the temperature sensors along the jacks. The temperature changes $\Delta T$
are calculated with respect to a reference temperature at $17^{\circ}$C. \\
Because of the spindle thread, a jack screw length change can occur which is not detected by the encoder on the 
worm gear shaft. Using Euler angles to calculate the kinematics of each jack screw, this undetected rotation $\beta_i$ can be
expressed with respect to a reference angle $\beta_{ref,i}$, yielding a rotation error compensation $\Delta L_{r,i}$ of the form:

\begin{equation}
\Delta L_{r,i}=(\beta_{ref,i}-\beta_i)\frac{p}{2\pi},
\end{equation}
where p=20 mm/rotation is the jack screw pitch.

A support cone compensation model takes into account the deformation of the cone due to temperature changes. The slight shift
of the lower fixed u-joints is then calculated by assuming that they expand on concentric circles with a temperature
averaged over several sensors in the cone:
\begin{eqnarray}
x_{new}&=&x+x\,\alpha\,(T-T_0),\\
y_{new}&=&y+y\,\alpha\,(T-T_0),
\end{eqnarray} 
where $x$ and $y$ are measured reference coordinates of the lower u-joints at a reference temperature $T_0=17^{\circ}$C.
$\alpha=12.0 \times 10^{-6}$K$^{-1}$ as for the jack screws.
When calculating the required jack lengths from the requested telescope position, this coordinate change results 
in a slight change of the required jack lengths.

Besides the group of jack screw corrections, a second group of telescope corrections is implemented: {\it radio and optical
refraction} (e.g. \citet{patel}),
 an {\it optical telescope (OT) collimation error correction} and an {\it interpolation table (IT)} for residual errors.
The radio refraction algorithm for the elevation correction $\Delta El_{ref,radio}$ in radians
is taken from \citet{allen73}:   
\begin{equation}
\Delta El_{ref,radio}=ref_0/\tan(El),
\end{equation}
where $ref_0=(N^2-1)/2N^2$ with $N=1-(7.8\times 10^{-5}\times P+0.39\times w_v/T)/T$. 
$T$,$P$ and $w_v$ are the temperature in K, the atmospheric 
pressure in mbar and the water vapour pressure in mbar, respectively. The calculation of the water vapour pressure $w_v$ is based
on the measured relative humidity $H$ on the site:
\begin{displaymath}
w_v=\frac{H}{100}P_{sat},   \nonumber
\end{displaymath}
where the saturation pressure of the water vapour in mbar is $P_{sat}=c_0\times 10^{c_1\times T/(c_2+T)}$ 
with the numerical 
constants $c_0=6.1078$, $c_1=7.5$, $c_2=237.3$ and the temperature in $^{\circ}$C.\\
The optical refraction correction in radians used for optical pointing observations is 
 adopted from \citet{seidelmann92} and 
implemented as:
\begin{equation}
\Delta El_{ref,opt}  =  1.2\times \frac{P}{1013.2}\frac{283.15}{T} 
  \times \frac{60.101\tan(d_Z)-0.0668\tan^3(d_Z)}{(180/\pi)3600} 
\end{equation}
where $d_Z$, $P$ and $T$ are the distance from the zenith in radians, the atmospheric pressure in mbar and the 
ambient temperature in K, respectively.

For both refraction correction modes we use annually-averaged values for the weather data. Extreme weather variations 
cause changes
in the refraction corrections of the order of fractions of arcseconds and are therefore negligible for our wavelength and 
antenna size. 

The optical telescope (OT) collimation error is corrected in the form:
\begin{eqnarray}
\Delta Az_{OT} &=&\frac{H_x\cos(\phi_{az}+\phi_{pol})+H_y\sin(\phi_{az}+\phi_{pol})}{\cos(\phi_{el})}\label{collimation1}\\
\Delta El_{OT} &=& H_y\cos(\phi_{az}+\phi_{pol})-H_x\sin(\phi_{az}+\phi_{pol}), \label{collimation2} 
\end{eqnarray} 
where $H_x$ and $H_y$ are the two tilt angles of the OT with respect to the mount pointing axes 
in the reference plane of the platform. $\phi_{az}$ and $\phi_{pol}$ are the mount azimuth and the platform
polarization, respectively.

A final pointing correction is done with an interpolation table (IT). Small measured pointing errors which 
are not explicitly modeled can be entered here. The three-dimensional table yields corrections 
$\Delta Az_{IT}, \Delta El_{IT}$
 as a function of $Az$, $El$ and $hexpol$ through a linear interpolation. 
Typically, we build the table with about 100 entries, the largest discontinuities of $\sim 50^{\circ}$ are
at the highest elevations around $85^{\circ}$. Lower elevation with larger errors are resolved at  $\sim 20^{\circ}$.

For the control system 
the sums of jack and telescope corrections are then used:
\begin{eqnarray}
\Delta L_{tot,i}&=&\Delta L_{p,i}+\Delta L_{T,i}+\Delta L_{r,i},\label{L_tot}\\
\Delta Az_{tot}&=&\Delta Az_{IT}+\Delta Az_{OT},\\
\Delta El_{tot}&=&\Delta El_{IT}+\Delta El_{OT}+\Delta El_{ref},\label{El_tot}
\end{eqnarray}
where $\Delta Az_{tot}$ and  $\Delta El_{tot}$ again lead to an effective change in jack length which is
subject to pointing corrections. $\Delta El_{ref}$ is either the radio or optical refraction correction.

All the pointing corrections are calculated on the PTC (Pointing Computer). 
From here they are transferred to the ACU in the
form described in equations(\ref{L_tot})-(\ref{El_tot}), illustrated in the flow chart in Figure \ref{fig5}.
The ACU applies corrections 
as:
\begin{equation}
L_{act,i}=L_{enc,i}+\Delta L_{tot,i},  \label{lact}
\end{equation}
where $L_{enc,i}$ and $L_{act,i}$ are the encoder measured and the real actual length of jack $i$, respectively. 
Equally, the actual position $(Az_{act},El_{act})$, defined as the real position 
after applying all the pointing corrections, is 
displayed on ACU and reported to the remote  TCS (Telescope Control System) as:
\begin{eqnarray}
Az_{act}&=&\widetilde{Az}+\Delta Az_{tot},\\
El_{act}&=&\widetilde{El}+\Delta El_{tot},
\end{eqnarray}
where $(\widetilde{Az},\widetilde{El})$  
is the telescope position as calculated from the corrected jack lengths $L_{act,i}$ in equation(\ref{lact}). 
The block diagram for the local control system is shown in Figure \ref{fig6}.

\section{hexpol, skypol and the sky coordinates} \label{appendix_angle}

\begin{figure}
\begin{center}
\includegraphics[scale=0.50,angle=0]{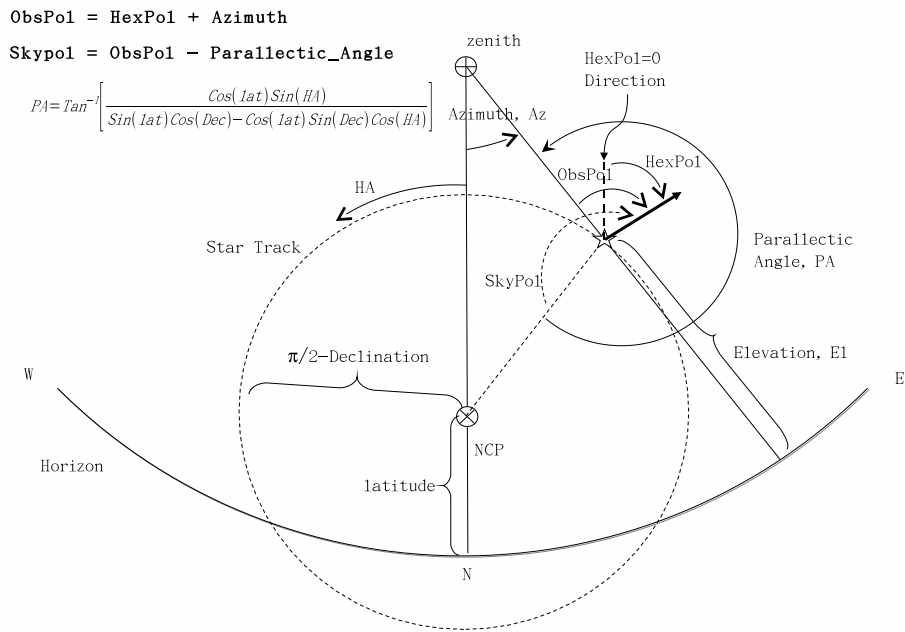}
\caption{\label{fig20}
The fundamental angles in astronomy and their relations in the mount and the celestial coordinate systems.}
\end{center}
\end{figure}

The interferometer baselines are fixed with respect to the platform. To describe the 
platform orientation, we use the mount coordinate system ($Az$, $El$, $ObsPol$), where $ObsPol$ refers to the angle 
between a specified axis on the platform and the line joining the zenith and the current pointing. 
This specified axis points toward south when the hexapod is in neutral position. 
The platform rotation is nominally related to the pointing by the relation $ObsPol=Az$. However, 
the hexapod is capable of changing the rotation by $\pm 30^{\circ}$ away from this relation. 
We define the added rotation as $hexpol$ and rewrite the relation as $ObsPol=Az+hexpol$.\\
The platform orientation can also be written in the sky coordinates ($ha$, $dec$, $skypol$), 
where $ha$ and $ra$ are the  hour angle and declination, respectively, in the equatorial coordinate system.
$skypol$ thus describes the angle between the specified axis and the line joining the current pointing and the 
North Celestial Pole (NCP). The transformation between the two system depends on the latitude ($lat$) of the 
observatory as  written below.
%Table \ref{tab:AnglesInAMiBA} summarizes what each angle means. 
Figure \ref{fig20} illustrates the relation between the different angles and the two coordinate systems.
%Specific to AMiBA, the definition of platform orientation (and hence the baseline orientation) requires particular explanation.
The basic transformation relations, including parallactic angle ($PA$), are:
\begin{eqnarray}
	\sin(El)&=&\sin(lat)\sin(dec)
                  +\cos(lat)\cos(dec)\cos(ha),\nonumber \\
	\cos(El)\cos(Az)&=&\cos(lat)\sin(dec)
                         -\sin(lat)\cos(dec)\cos(ha),\nonumber \\
	\cos(El)\sin(Az)&=&-\cos(dec)\sin(ha), \nonumber \\
	skypol&=& obspol-PA,\nonumber \\
	PA &=& 
\tan^{-1}\left(\frac{\cos(lat)\sin(ha)} {\sin(lat)\cos(dec) - \cos(lat)\sin(dec)cos(ha)}\right).\nonumber
\end{eqnarray}

\end{document}